\pdfoutput=1
\documentclass[11pt,a4paper]{article}
\usepackage[margin=25mm]{geometry}
\usepackage{booktabs}
\usepackage{microtype}
\usepackage{graphicx}
\usepackage{multicol,multirow}
\usepackage{amsmath,amssymb,amsfonts}
\usepackage{mathrsfs}
\usepackage{amsthm}
\usepackage{rotating}
\usepackage{appendix}
\usepackage[authoryear, round]{natbib}
\usepackage{ifpdf}
\usepackage[T1]{fontenc}%
\usepackage{times}%
\usepackage{newtxmath}%
\usepackage{textcomp}%
\usepackage{xcolor}%
\usepackage{bm}
\usepackage{tikz}
\usetikzlibrary{arrows.meta, positioning, calc, backgrounds}
\usepackage{pgfplots}
\pgfplotsset{compat=1.18}
\usepackage{hyperref}
\usepackage{algorithm}
\usepackage{algpseudocode}
\algnewcommand\algorithmicinput{{\bfseries Input:}}
\algnewcommand\Input{\item[\algorithmicinput]}
\algnewcommand\algorithmicoutput{{\bfseries Output:}}
\algnewcommand\Output{\item[\algorithmicoutput]}

\makeatletter
\newcommand{\algparbox}{\parbox[t]{\dimexpr\linewidth-\ALG@thistlm}}
\makeatother

\hypersetup{colorlinks=true, citecolor=teal, linkcolor=teal, urlcolor=teal}
\title{Forward-Backward Map Matching for Airborne Magnetic Anomaly Navigation}
\author{Joseph O'Leary\textsuperscript{1,2}\thanks{Corresponding author: \href{mailto:jdoleary@swin.edu.au}{jdoleary@swin.edu.au}} \and
Tapio Simula\textsuperscript{3} \and
Tom Kimpson\textsuperscript{4,5} \and
Corey Davey\textsuperscript{1} \and
Wenchao Li\textsuperscript{6} \and
Andrew D. Greentree\textsuperscript{6} \and
Allison Kealy\textsuperscript{1}}
\date{}

\begin{document}

\maketitle

\begin{center}
\small
\textsuperscript{1}Innovative Planet Research Institute, Swinburne University of Technology, Melbourne, Australia\\
\textsuperscript{2}School of Physics, University of Melbourne, Melbourne, Australia\\
\textsuperscript{3}Optical Sciences Centre, Swinburne University of Technology, Melbourne, Australia\\
\textsuperscript{4}School of Mathematics and Statistics, The University of Melbourne, Melbourne, Australia\\
\textsuperscript{5}ARC Centre of Excellence for the Mathematical Analysis of Cellular Systems (MACSYS), The University of Melbourne, Melbourne, Australia\\
\textsuperscript{6}Department of Physics, School of Science, RMIT University, Melbourne, Australia
\end{center}

\begin{abstract}
Airborne magnetic anomaly navigation provides a passive, unjammable alternative to Global Navigation Satellite Systems positioning by matching magnetometer measurements against a lithospheric magnetic anomaly map. Existing map-matching algorithms either assume a Gaussian posterior (extended Kalman filter), require Monte Carlo sampling (particle filter), or discard posterior uncertainty in favour of a single optimal path (Viterbi algorithm). Here we formulate map matching as inference on a chain-structured factor graph and solve it segment by segment using the forward-backward algorithm, which yields a discrete posterior over candidate positions and its Shannon entropy $H_t$ at time $t$. Over the \texttt{Renfrew\_395} map, $10^3$ synthetic trials yield a median position error of $32.7$ m, $91\%$ below the median inertial drift of $362$ m. On these trials and $10^3$ flight paths, the median error is $43\%$ below the extended Kalman filter and particle filter. Across these paths, a threshold on segment-level entropy detects $47\%$ of segments with errors above two map cells and $94\%$ of those above three, at a $10\%$ false-alarm rate. The foregoing results indicate that $H_t$ provides a truth-free diagnostic of map-conditioned navigation ambiguity.
\end{abstract}

\section{Introduction}\label{Sec:Intro} 
 
Global Navigation Satellite Systems [GNSS; \cite{Langley_2017}] such as the Global Positioning System underpin positioning, navigation, and timing capabilities of worldwide critical infrastructure, e.g.\ telecommunications, financial markets, precision agriculture, and aviation \citep{Kaplan_2017}. Because GNSS signals are weak ($\sim -10^2 \, \rm{dBW}$) when received at Earth \citep{Kealy_2017}, they are susceptible to jamming, spoofing, and denial in contested or obstructed environments \citep{Psiaki_2016,Rados_2024}. This growing vulnerability of GNSS signals to interference has prompted searches for alternative navigation modalities that can bound the drift of inertial navigation systems [INS; \cite{Woodman_2007}] in GNSS-degraded environments \citep{Groves_2013}. 
 
For airborne platforms, one possible alternative to satellite-based navigation is magnetic anomaly navigation [MagNav; \cite{Canciani_2016,Canciani_2017}]. In its simplest form, MagNav exploits the spatial variations in the Earth's lithospheric (``crustal'') magnetic field by comparing onboard magnetometer measurements against presurveyed magnetic anomaly maps; see Chapters 11--13 of \cite{Hinze_2013} for summaries about magnetic data acquisition, processing, and interpretation, respectively. The lithospheric field is globally available and time-invariant on navigation timescales, i.e.\ $\lesssim 10^5 \, \rm{s}$, with root-mean-square secular variations $\sim 10^2 \, \rm{pT \, yr^{-1}}$ \citep{Thebault_2009,Meyer_2017}. Accordingly, MagNav has emerged as a promising navigation modality, providing a passive, map-referenced source of positioning information that is resilient to  jamming and spoofing, with practical demonstrations for airborne estimation \citep{Canciani_2016,Canciani_2017}, simultaneous localisation and mapping \citep{Lee_2020}, and flight-test studies \citep{Canciani_2021,muradoglu_2025}. 

There are at least three separate challenges associated with map-matching approaches to MagNav. First, the inference problem can be ill-conditioned or multimodal in regions where the anomaly field varies weakly or is spatially repetitive, giving rise to multiple candidate trajectories consistent with the measurements \citep{Wang_2023a,Wang_2023b}. Second, the lithospheric field must be isolated from other contributions measured by the onboard magnetometer, e.g.\ the core geomagnetic field, diurnal ionospheric variations,  as well as platform-generated interference \citep{Gnadt_2022b,Lathrop_2024,Hager_2026}. Third, the inference algorithm must balance computational efficiency against the fidelity with which it represents the posterior distribution over candidate positions. Two popular estimation algorithms in MagNav applications, namely the extended Kalman filter [EKF; \cite{Canciani_2017}] and particle filter \citep{Canciani_2016,Lee_2020}, represent complementary points in this trade-off space. For example, the EKF is computationally efficient and widely deployed, but assumes that the posterior is unimodal and Gaussian. The particle filter captures multimodality through weighted samples, but requires tuning to avoid degeneracy. Both are causal filters that do not revisit earlier estimates when informative measurements arrive later in the trajectory. 
 
This paper addresses the third challenge by formulating MagNav map-matching as inference on a chain-structured factor graph \citep{Kschischang_2001, Loeliger_2004} and solving it recursively using the forward-backward algorithm \citep{Rabiner_1989,Quinn_2001}. The method divides the flight into segments of equal length and discretises the aircraft position onto a map-aligned grid of candidate cells. Within each segment, the forward sweep propagates the probability of each cell along the displacement reported by the onboard INS and weights it by the likelihood of each new magnetometer measurement. Once the last measurement of the segment arrives, the backward sweep conditions the probability at each timestep on the later measurements. Taken together, the forward and backward sweeps yield the marginal posterior over the grid cells at each timestep, from which we compute an aircraft position estimate and the Shannon entropy of the posterior \citep{Shannon_1948}. The results in this paper build on previous works on the point-mass filter [PMF; \cite{Bucy_1971}; Section~2.3 of \cite{Bergman_1999}], whose recursion is the forward sweep of the forward-backward algorithm and which has been applied to terrain-aided navigation \citep{Bergman_1999b}, and on grid-based smoothing \citep{Kitagawa_1987,Dunik_2019}. The formulation extends the hidden Markov model approach of \cite{Li_2024}, who apply the classic Viterbi algorithm \citep{Viterbi_1967,Rabiner_1989} to gravity-aided navigation, from maximum a posteriori (MAP) sequence estimation to full marginal posterior inference via the sum-product algorithm. More broadly, factor graphs have found wide application in navigation, including simultaneous localisation and mapping \citep{Dellaert_2006}, GNSS/INS integration \citep{Wen_2021}, terrain-referenced navigation \citep{Park_2024}, and magnetic sensor compensation \citep{Lathrop_2024}.

The factor graph formulation adopted in this paper complements the aforementioned navigation algorithms, i.e.\ the EKF, particle filter, and Viterbi algorithm, in at least four ways: (i) the posterior is represented explicitly as a probability mass function over position grid cells, exposing multimodal ambiguity without sampling; (ii) the Shannon entropy is employed as a closed-form, truth-free diagnostic of posterior uncertainty; (iii) unlike causal filters, the factor graph revises each estimate with the later measurements of its segment, at the cost of a delay of up to one segment; and (iv) the forward-backward position estimate is a posterior-weighted average over the candidate grid, i.e.\ the estimate is not constrained to grid cell centres, unlike the MAP path returned by the Viterbi algorithm. We validate the factor graph formulation using synthetic INS and scalar magnetometer data generated over a real aeromagnetic anomaly map.

The remainder of this paper is structured as follows. In Section~\ref{Sec:FG} we discuss inference on a chain-structured factor graph using the forward-backward algorithm \citep{Rabiner_1989,Quinn_2001}. We specialise the factor graph of Section~\ref{Sec:FG} to magnetic anomaly navigation in Section~\ref{Sec:Magnav}. In Section~\ref{Sec:Validation} we present a worked example of airborne magnetic anomaly navigation using the factor graph formalism of Sections \ref{Sec:FG} and \ref{Sec:Magnav}. Monte Carlo realisations of an airborne INS are performed in Section~\ref{Sec:Performance}, where the factor graph is compared with the PMF, the EKF, the particle filter, and the Viterbi path, and a threshold on the posterior entropy is tested as a flag for large position errors. Some practical implications for airborne magnetic anomaly navigation are canvassed briefly in Section~\ref{Sec:Conclusion}.

\section{Factor graph formalism}\label{Sec:FG}

A factor graph on a chain \citep{Kschischang_2001, Loeliger_2004} provides an efficient statistical framework for inferring the evolution of a hidden state from a time-ordered sequence of observations \citep{Dellaert_2006,Thrun_2006}. In the MagNav context, the hidden state is the aircraft position on a discrete map-aligned grid, the observations are magnetic anomaly measurements, and the factor graph is solved recursively for the hidden state marginal posterior distribution using the forward-backward algorithm \citep{Rabiner_1989,Quinn_2001}. In Section~\ref{SubSec:FGChain} we introduce the general chain-structured formulation. Forward-backward inference, minimum mean square error (MMSE) point estimation, and the Shannon entropy of the marginal posterior are summarised in Sections \ref{SubSec:forward_backward} and \ref{SubSec:point_estimation}, respectively. 

\subsection{Factor graph on a chain}\label{SubSec:FGChain}
\begin{figure}[t]
\hspace{0.5cm}
\includegraphics{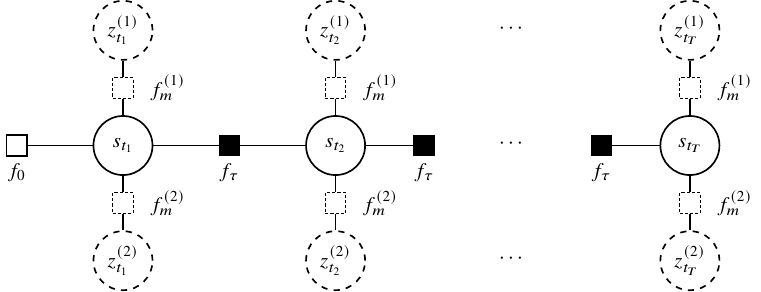}
\caption{
A chain-structured factor graph with prior $f_0$, measurement $f_m^{(\ell)}$, and transition $f_{\tau}$ factors. Hidden states $s_{t_1}, \ldots, s_{t_T}$ (solid circles)
are connected by transition factors $f_\tau$ (filled squares). Two
families of measurement factors, $f_m^{(1)}$ and $f_m^{(2)}$ (dashed
open squares), link each hidden state to corresponding observations
$z_t^{(1)}$ and $z_t^{(2)}$ (dashed circles). A prior factor $f_0$
(open square with solid border) is attached to the initial state}
\label{fig:chain_factor_graph_multi}
\end{figure}

At time $t \in \mathcal{T} = \{t_1, \dots ,t_{T} \}$, the system occupies a hidden state $s_t \in \mathcal{S}_t = \{c_{t}^{1}, \dots, c_{t}^{N_{s}}\}$, where $\mathcal{S}_t$ is the set of $N_s$ candidate states and $c^{j}_{t}$ denotes the $j$-th candidate state at time $t$, for $j = 1, \dots, N_s$. Let $z_t$ denote the observation at time $t$. The posterior distribution over the full state sequence is (up to a normalisation constant) 
\begin{equation}
p(s_{\mathcal{T}} \mid z_{\mathcal{T}})
\propto
f_0(s_{t_1})
\prod_{t \in \mathcal{T}} f_m(z_t, s_t)
\prod_{k=2}^{T} f_\tau(s_{t_k},s_{t_{k-1}}),
\label{eq:posterior_prop}
\end{equation}
where $s_{\mathcal{T}} = (s_{t_1}, \ldots, s_{t_T})$ and
$z_{\mathcal{T}} = (z_{t_1}, \ldots, z_{t_T})$. In this paper, we express Equation~(\ref{eq:posterior_prop}) as a factor graph with chain topology whose factors correspond to a first-order hidden Markov model, viz.\
$f_0(s_{t_1}) = p(s_{t_1})$, $f_m(z_{t_{k}}, s_{t_{k}}) = p(z_{t_{k}} \mid s_{t_{k}})$, and $f_\tau(s_{t_k},s_{t_{k-1}}) = p(s_{t_k} \mid s_{t_{k-1}})$; see \cite{Rabiner_1989} for an overview of the theory of hidden Markov models. That is, the components $f_0$, $f_m$, and $f_\tau$ of Equation~(\ref{eq:posterior_prop}) denote the prior, measurement, and transition factors, respectively. A general factor graph with chain topology is depicted schematically in Figure~\ref{fig:chain_factor_graph_multi}.\footnote{Although the present work considers a single sensor, each additional sensor contributes its own measurement factor attached to the same hidden state $s_t$, and the measurement factor in Equation~(\ref{eq:posterior_prop}) is replaced with $\prod_{\ell = 1}^{L} f_m^{(\ell)}[z^{(\ell)}_t, s_t]$. The conditional independence of the sensor measurements given the hidden state is encoded in the graph topology.} In practice, the choice of these factors is specific to the problem and lies with the analyst. They are specialised to the problem of MagNav in Section~\ref{Sec:Magnav} and we refer the reader to \cite{Taylor_2024} for further details about applying factor graphs to problems related to navigation.

\subsection{Forward-backward inference}
\label{SubSec:forward_backward}
\begin{figure}[t]
\centering
\includegraphics{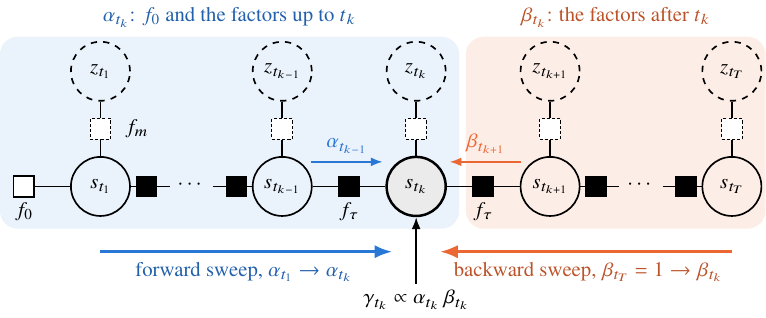}
\caption{
Forward and backward sweeps on the chain of Figure~\ref{fig:chain_factor_graph_multi}, with one measurement factor per timestep. The forward variable $\alpha_{t_k}$ collects the prior factor $f_0$ and the factors up to $t_k$ (blue), and the backward variable $\beta_{t_k}$ collects the factors after $t_k$ (orange). The forward sweep computes $\alpha_{t_1}, \dots, \alpha_{t_k}$ by Equations~(\ref{eq:alpha_init}) and~(\ref{eq:alpha_rec}), and the backward sweep computes $\beta_{t_T}, \dots, \beta_{t_k}$ by Equations~(\ref{eq:beta_init}) and~(\ref{eq:beta_rec}). The normalised product of $\alpha_{t_k}$ and $\beta_{t_k}$ at $s_{t_k}$ (shaded) is the marginal posterior $\gamma_{t_k}$ of Equation~(\ref{eq:gamma})}
\label{fig:forward_backward}
\end{figure}

Summing Equation~(\ref{eq:posterior_prop}) over the states at all timesteps except $t_k$ gives the marginal posterior at node $s_{t_k} \in \{ c_{t_k}^{1}, \dots , c_{t_k}^{N_s}\}$ of the chain in Figure~\ref{fig:chain_factor_graph_multi},
\begin{equation}
p(s_{t_k} \mid z_{\mathcal T})
\propto
\sum_{s_{\mathcal T}\setminus s_{t_k}} \left[
f_0(s_{t_1})
\prod_{r=1}^{T} f_m(z_{t_r},s_{t_r})
\prod_{r=2}^{T} f_\tau(s_{t_r},s_{t_{r-1}})\right],
\label{eq:marginal_factorised}
\end{equation}
where $s_{\mathcal T}\setminus s_{t_k}$ denotes the states summed over. For a fixed candidate value $s_{t_k}=c_{t_k}^j$, Equation~(\ref{eq:marginal_factorised}) contains $N_s^{T-1}$ complete hidden-state sequences consistent with $s_{t_k}=c_{t_k}^j$ and is therefore computationally prohibitive for most practical problems. Forward-backward inference circumvents this by exploiting the conditional independence encoded by the chain topology. For a fixed node $s_{t_k}$, Equation~(\ref{eq:marginal_factorised}) may be grouped into a left contribution involving the prior factor $f_0$, the transition factors $f_{\tau}$, and the measurement factors $f_{m}$ up to time $t_k$, and a right contribution involving the transition and measurement factors from time $t_{k+1}$ onward. The left and right contributions are summarised recursively by the forward and backward variables, denoted by $\alpha$ and $\beta$, respectively. The forward and backward sweeps that compute $\alpha$ and $\beta$ are depicted schematically in Figure~\ref{fig:forward_backward}. At node $s_{t_k}$, the normalised product of $\alpha_{t_k}$ and $\beta_{t_k}$ is the marginal posterior $\gamma_{t_k}$ of Equation~(\ref{eq:gamma}). This dynamic programming \citep{Bellman_1957,Rabiner_1989,Quinn_2001} structure replaces explicit summation over complete state sequences with a recursion whose cost scales as $O(TN_s^2)$ (the quadratic factor arises because each of the $N_s$ candidate states at time $t_k$ requires a sum over all $N_s$ predecessor states at $t_{k-1}$). This cost may be reduced further via active-set pruning (Section~\ref{SubSec:Pruning}), which restricts each recursion step to a subset $\mathcal{A}_{t_k} \subset \mathcal{S}_{t_k}$ of candidate states with non-negligible posterior mass, replacing $N_s^2$ with $|\mathcal{A}_{t_k}|^2$ per timestep.

The marginal posterior $\gamma_{t_{k}}(j)$ of candidate $c_{t_{k}}^j$ at time $t_k$ is given by
\begin{equation}
\gamma_{t_{k}}(j)
=
\frac{\alpha_{t_k}[j]\beta_{t_k}[j]}
{\sum_{\ell=1}^{N_s}\alpha_{t_k}[\ell]\beta_{t_k}[\ell]}, 
\label{eq:gamma}
\end{equation}
where $\gamma_{t_k}(j) = p(s_{t_k} = c_{t_k}^j \mid z_{\mathcal{T}})$. The forward variable $\alpha_{t_k}[j] = p(s_{t_k} = c_{t_k}^j,\; z_{t_1:t_k})$ is the joint probability of candidate state $c_{t_k}^j$ and the observations $z_{t_1:t_k} = (z_{t_1}, \dots, z_{t_k})$. The backward variable $\beta_{t_k}[j] = p(z_{t_{k+1}:t_T} \mid s_{t_k} = c_{t_k}^j)$ is the likelihood of the future observations $z_{t_{k+1}:t_T} = (z_{t_{k+1}}, \dots, z_{t_T})$ conditioned on $s_{t_k} = c_{t_k}^j$. 
 
The forward and backward recursions follow directly from the chain structure. For the forward variable, the initial condition is
\begin{equation}
\alpha_{t_1}[j]
=
f_0(c_{t_1}^j)\,f_m(z_{t_1},c_{t_1}^j),
\label{eq:alpha_init}
\end{equation}
and for $k=2,\dots,T$,
\begin{equation}
\alpha_{t_k}[j]
=
f_m(z_{t_k},c_{t_k}^j)
\sum_{i=1}^{N_s}
\alpha_{t_{k-1}}[i]\,
f_\tau(c_{t_k}^j,c_{t_{k-1}}^i).
\label{eq:alpha_rec}
\end{equation}
For the backward variable, the terminal condition is
\begin{equation}
\beta_{t_T}[j]=1,
\label{eq:beta_init}
\end{equation}
and for $k=T-1,\dots,1$,
\begin{equation}
\beta_{t_k}[j]
=
\sum_{\ell=1}^{N_s}
f_\tau(c_{t_{k+1}}^\ell,c_{t_k}^j)\,
f_m(z_{t_{k+1}},c_{t_{k+1}}^\ell)\,
\beta_{t_{k+1}}[\ell].
\label{eq:beta_rec}
\end{equation}
In practice, the forward and backward variables are renormalised at each timestep to avoid numerical underflow, the practical details of which are discussed in \cite{Rabiner_1989}.

Three properties of the forward-backward algorithm carry over to the magnetic anomaly navigation of Section~\ref{Sec:Magnav}. (i) One forward sweep and one backward sweep give $\gamma_{t_k}$ at every timestep of the chain, which in the application is one segment of $T$ timesteps (Section~\ref{SubSec:Segmentation}). (ii) The forward sweep is the PMF \citep[Section~2.3]{Bergman_1999}, and $\alpha_{t_k}$, normalised over the candidates, is the filtering density $p(s_{t_k} \mid z_{t_1:t_k})$. The backward sweep incorporates the later measurements $z_{t_{k+1}:t_T}$ through $\beta_{t_k}$, and its contribution is quantified in Section~\ref{SubSec:Comparison}. (iii) The position estimate and the entropy diagnostic of Section~\ref{SubSec:point_estimation} are read off $\gamma_{t_k}$.

\subsection{Point estimation and posterior diagnostics}
\label{SubSec:point_estimation}

Given the marginal posterior $\gamma_{t_k}(j)$ of Section~\ref{SubSec:forward_backward}, the MMSE position estimate $\hat{s}_{t_k}^{\rm MMSE}$ at time $t_k$ is
\begin{equation}
    \hat{s}_{t_k}^{\rm MMSE} = \sum_{j=1}^{N_s} \gamma_{t_k}(j)\, c^j_{t_k},
    \label{eq:mmse}
\end{equation}
and the marginal MAP estimate is the candidate of highest marginal posterior,
\begin{equation}
    \hat{s}_{t_k}^{\rm MAP} = c_{t_k}^{j^*},
    \label{eq:map}
\end{equation}
where $j^* = \arg\max_j \gamma_{t_k}(j)$. When the posterior is degenerate, i.e.\ $\gamma_{t_k}(j^*) = 1$, Equations~(\ref{eq:mmse}) and~(\ref{eq:map}) coincide, viz.\ $\hat{s}_{t_k}^{\rm MMSE} = \hat{s}_{t_k}^{\rm MAP} = c^{j^*}_{t_k}$. When the posterior is uniform, i.e.\ $\gamma_{t_k}(j) = N_s^{-1}$, the MMSE estimate reduces to the centroid of the candidate grid.

The MMSE estimate uses the full posterior and can lie between cell centres, while the marginal MAP estimate lies on a cell centre. Given a posterior with well-separated modes, the MMSE estimate can fall between the modes, at a position to which the posterior assigns low probability. In this case, a mode-tracking estimate, such as the marginal MAP estimate of Equation~(\ref{eq:map}) or the Viterbi path, is preferable. The Viterbi algorithm \citep{Viterbi_1967,Rabiner_1989} returns the most probable complete sequence, $\hat{s}^{\rm Viterbi}_{\mathcal T} = \arg\max_{s_{\mathcal T}} p(s_{\mathcal T} \mid z_{\mathcal T})$. It runs on the same chain as the forward-backward algorithm, with the sum in the forward recursion of Equation~(\ref{eq:alpha_rec}) replaced by a maximum, followed by a traceback step that recovers the path. For the convenience of the reader, the Viterbi recursion is written out in full in Appendix~\ref{App:Viterbi}. The cells that Equation~(\ref{eq:map}) selects one timestep at a time do not in general form the Viterbi path. Looking ahead to the results of Section~\ref{SubSec:Comparison}, the MMSE estimate of Equation~(\ref{eq:mmse}) yields a lower median position error than the marginal MAP estimate and the Viterbi path.

We employ the Shannon entropy $H_t$ of the marginal posterior \citep{Shannon_1948},
\begin{equation}
    H_t = -\sum_{j=1}^{N_s} \gamma_t(j) \ln \gamma_t(j),
    \label{eq:entropy}
\end{equation}
as a scalar diagnostic of posterior concentration. The Shannon entropy satisfies $H_t \in [0, \ln N_s]$, with $H_t = 0$ corresponding to the degenerate case and $H_t = \ln N_s$ to the uniform case. Because it is calculated from $\gamma_t$, $H_t$ is available without access to the true trajectory. In an operational setting where the true trajectory is unknown, $H_t$ provides the navigator with a per-timestep indicator of whether the map-matching solution is well-constrained or ambiguous.

The value of $H_t$ in nats depends on the grid design, through the number of candidates $N_s$, the grid extent, the measurement factor $f_m$, and the pruning of Section~\ref{SubSec:Pruning}, so a threshold on $H_t$ is specific to one grid design. We report alongside $H_t$ the normalised entropy
\begin{equation}
    \bar{H}_t = \frac{H_t}{\ln N_s},
    \label{eq:entropy_norm}
\end{equation}
which lies between 0 and 1 for any $N_s$. With pruning, $\gamma_t$ is computed on the active set $\mathcal{A}_t$, so the pruned cells carry zero mass and $H_t$ is biased low. For the interested reader, Appendix~\ref{App:ActiveSetPruning} quantifies this bias as a function of the pruning threshold, together with the computational speed-up and the change in position error.

In this section, the hidden states $s_t$, candidate states $c^j_t$, and observations $z_t$ are written without reference to their dimensionality. The forward-backward recursion and its associated entropy diagnostic $H_t$ operate on the candidate index $j$ and hence are agnostic to whether $c^j_t$ is a scalar or vector. 

\section{Magnetic anomaly navigation}
\label{Sec:Magnav}
The lithospheric magnetic anomaly field varies spatially on scales $\lesssim 10^3 \, \rm{km}$ with anomaly amplitudes $\lesssim 10^3 \, \rm{nT}$ at airborne survey altitudes $\lesssim 5 \, \rm{km}$, several orders of magnitude above the measurement accuracy ($\sim 1 \, \rm{nT}$) of optically pumped magnetometers \citep{Hinze_2013,Thebault_2015,Canciani_2017}. In this section we
specialise the prior $f_0$, transition $f_{\tau}$, and measurement $f_{m}$
factors of Section~\ref{Sec:FG} to the problem of MagNav. The problem
formulation is introduced in Section~\ref{SubSec:ProbForm}. A recipe for
discretising the aircraft position domain onto a map-aligned grid is given
in Section~\ref{SubSec:Discretisation}. The factors $f_0$, $f_{\tau}$, and
$f_{m}$ are defined in Sections~\ref{SubSec:Prior}, \ref{SubSec:Trans}, and
\ref{SubSec:Meas}, respectively. Some simplifying model idealisations associated with the foregoing factors are mentioned briefly in Section~\ref{SubSec:Idealisations}. Flight-time segmentation, segment-length
selection, and active-set pruning are
discussed in Sections~\ref{SubSec:Segmentation}--\ref{SubSec:Pruning}.

\subsection{Problem formulation}
\label{SubSec:ProbForm}

Here we consider an aircraft equipped with a scalar magnetometer and an INS flying over a region for which a reference magnetic anomaly map $B_{\rm anom}(\phi, \lambda)$ is available. The map is defined on a regular grid in geodetic coordinates $(\phi, \lambda)$ and assigns to each grid point a scalar anomaly value, obtained from aeromagnetic survey data after removal of the core field and external variations \citep{Hinze_2013}. Values at arbitrary positions are obtained by interpolation.

At each timestep $t_k$ for $k = 1, \ldots, N$, the aircraft occupies an unknown true position $\bm{s}_{t_{k}} \in \mathbb{R}^2$ and the following data are available: (i) a scalar magnetic anomaly measurement $z_{t_{k}} \in \mathbb{R}$; (ii) an INS velocity estimate $\bm{v}_{t_{k}} = [v^{\rm (N)}_{t_{k}}, v^{\rm (E)}_{t_{k}}]$ with north (N) and east (E) components; and (iii) an INS position estimate $\hat{\bm{s}}_{t_{k}}^{\rm INS} = (\hat{\phi}_{t_{k}}, \hat{\lambda}_{t_{k}}) \in \mathbb{R}^2$. The objective is to estimate the true flight path $\{\bm{s}_{t_{k}}\}_{k=1}^{N}$ by fusing the magnetic observations with the INS estimates via the factor graph of Section~\ref{Sec:FG}. The flight is processed in $M$ consecutive segments of $T$ timesteps each (see Section~\ref{SubSec:Segmentation}). The INS position estimate $\hat{\bm{s}}^{\rm INS}_{t_{k}}$ determines where the candidate grid is placed at each timestep (see Section~\ref{SubSec:Discretisation}), but does not enter the transition (see Section~\ref{SubSec:Trans}) or measurement (see Section~\ref{SubSec:Meas}) factors. The INS velocity $\bm{v}_{t_{k}}$ enters the transition factor. 

\subsection{State space discretisation}
\label{SubSec:Discretisation}
The factor graph of Section~\ref{Sec:FG} approximates the continuous position
$\bm{s}_{t_{k}}$ by restricting the hidden state to a finite candidate set
$\mathcal{S}_{t_{k}}$ on a map-aligned grid, i.e.\ an $n \times n$ subgrid of the survey grid on which the anomaly map is stored. The grid nodes are map nodes in geodetic coordinates, with rows along parallels and columns along meridians, and the grid is fixed to the Earth. At each timestep $t$ of segment $m$, the grid centre $\bm{s}^{(0)}_t = [\phi_t^{(0)}, \lambda_t^{(0)}]$ is the map node nearest to the INS position shifted by a carried correction,
\begin{equation}
\bm{s}^{(0)}_t = \mathcal{Q}\big[\hat{\bm{s}}^{\rm INS}_t + \Delta\hat{\bm{s}}^{(m-1)}\big],
\label{eq:grid_centre}
\end{equation}
where $t \in \mathcal{T}^{(m)}$, $\mathcal{Q}[\cdot]$ returns the nearest map node, and
\begin{equation}
\Delta\hat{\bm{s}}^{(m-1)} = \hat{\bm{s}}^{\rm MMSE}_{t_T^{(m-1)}} - \hat{\bm{s}}^{\rm INS}_{t_T^{(m-1)}},
\label{eq:carried_correction}
\end{equation}
with $\Delta\hat{\bm{s}}^{(0)} = \bm{0}$, is the offset between the MMSE estimate and the INS position at the last timestep of the preceding segment (see Algorithm~1 and Section~2 of \citealt{Li_2024} for further details). The candidate set
$\mathcal{S}_{t_{k}} = \{\bm{c}_{t_k}^1, \ldots, \bm{c}_{t_k}^{N_s}\}$
comprises $N_s = n \times n$ positions arranged symmetrically about
the grid centre, where $n$ is an odd integer
and each candidate
$\bm{c}_{t_k}^j = (\phi_{t_k}^j, \lambda_{t_k}^j)$ is a map node with map value $B_{\rm anom}(\bm{c}^{j}_{t_{k}})$. The candidate positions are
\begin{equation}
\phi_{t_k}^j = \phi_{t_k}^{(0)} + d_p\,\Delta_\phi,
\label{eq:candidate_offsets_phi}
\end{equation}
and
\begin{equation}
\lambda_{t_k}^j = \lambda_{t_k}^{(0)} + d_q\,\Delta_\lambda,
\label{eq:candidate_offsets_lambda}
\end{equation}
where $d_p, d_q \in \{-\lfloor n/2 \rfloor, \ldots, \lfloor n/2 \rfloor\}$
and $\Delta_{\phi}$, $\Delta_{\lambda}$ are the map grid spacings in latitude
and longitude, respectively. The numerical values of $\Delta_{\phi}$ and $\Delta_{\lambda}$ for the aeromagnetic anomaly map employed in this work are given in Section~\ref{SubSec:Scenario}.

With the grid recentred at every segment, the grid half-width $n_{\rm half}$, where
\begin{equation}
n = 2\,n_{\rm half} + 1,
\label{eq:grid_size_fixed_halfwidth}
\end{equation}
has to cover only the INS drift within one segment plus the error of the estimate at the preceding segment boundary (see Section~\ref{SubSec:SelectingT}). In the present application, we adopt $n_{\rm half} = 50$ and hence $n = 101$. This value is held fixed across all trials in Sections \ref{Sec:Validation} and \ref{Sec:Performance}, and Section~\ref{SubSec:SelectingT} uses it to bound the segment length $T$. All distance calculations below use the corresponding local north--east approximation to these geodetic coordinates.

\begin{figure}
\centering
\includegraphics{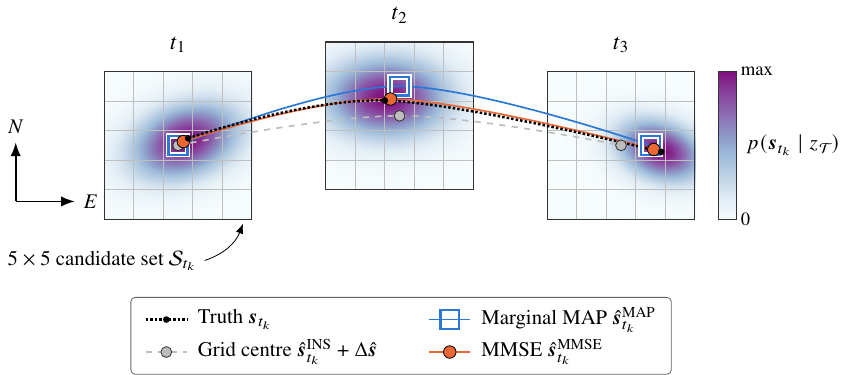}
\caption{Schematic of the candidate set $\mathcal{S}_{t_{k}}$ on a $5 \times 5$ grid at three successive timesteps $t_1$, $t_2$, and $t_3$. Each grid is centred on the map node nearest to $\hat{\bm{s}}^{\rm INS}_{t_k} + \Delta\hat{\bm{s}}$ (grey points, dashed curve), the INS position plus the correction carried from the previous segment [Equations~(\ref{eq:grid_centre}) and~(\ref{eq:carried_correction})]. The shading is the marginal posterior density $p(\bm{s}_{t_k} \mid z_{\mathcal{T}})$, and $\gamma_{t_{k}}(j)$ [Equation~(\ref{eq:gamma})] is its value at the cell centres, normalised over the grid. The marginal MAP estimate $\hat{\bm{s}}^{\rm MAP}_{t_k}$ [Equation~(\ref{eq:map}); blue squares, solid curve] is the cell of highest $\gamma_{t_k}$ at each timestep. The Viterbi path of Appendix~\ref{App:Viterbi}, the most probable sequence of cells over the whole segment, does not in general pass through these cells. The MMSE estimate $\hat{\bm{s}}^{\rm{MMSE}}_{t_{k}}$ [Equation~(\ref{eq:mmse}); orange points, solid curve] is the posterior-weighted average over the candidate set. The $5 \times 5$ grid is presented as a visual aid; the results in Sections~\ref{Sec:Validation} and \ref{Sec:Performance} adopt $n = 101$\label{Fig:Schematic}}
\end{figure}

Figure~\ref{Fig:Schematic} illustrates the candidate set on an $N_s = 5 \times 5$ grid at three successive timesteps $t_1$, $t_2$, and $t_3$. The candidate set $\mathcal{S}_{t_{k}}$ is constructed around the grid centre of Equation~(\ref{eq:grid_centre}) (grey points, dashed curve) at each timestep. The marginal MAP estimate $\hat{\bm{s}}^{\rm MAP}_{t_k}$ of Equation~(\ref{eq:map}) (blue squares, solid curve) is the cell of highest marginal posterior at each timestep. The Viterbi path of Appendix~\ref{App:Viterbi} is the most probable sequence of cells over the whole segment, and in general differs from the sequence of marginal MAP estimates. When the posterior mass is spread across several neighbouring cells along the true trajectory $\bm{s}_{t_{k}}$ (black points, dotted curve), the MMSE estimate $\hat{\bm{s}}_{t_{k}}^{\rm{MMSE}}$ of Equation~(\ref{eq:mmse}) (orange points, solid curve) can lie between cell centres, whereas the marginal MAP estimate and the Viterbi path are constrained to cell centres by construction. The $5 \times 5$ grid in Figure~\ref{Fig:Schematic} is illustrative only; the validation and performance tests of Sections~\ref{Sec:Validation} and \ref{Sec:Performance} adopt $n=101$. 

\subsection{Prior factor}
\label{SubSec:Prior}

The prior factor $f_0$ weights the candidate states at the first timestep of a
processing segment according to the available position information. Two cases
arise depending on whether the segment is the first in the flight or a
continuation of an earlier segment.
 
For the initial segment, no preceding posterior is available and the prior is
taken to be uniform over the candidate grid,
\begin{equation}
f_0\left(\bm{c}_{t_1}^j\right)
= \frac{1}{N_s},
\label{Eq:Prior_Magnav}
\end{equation}
for $j=1,\ldots,N_s$.

For each subsequent segment $m > 1$, the terminal marginal posterior
$\gamma_{t_T^{(m-1)}}$ from the preceding segment is propagated forward
through one application of the transition factor to yield a non-Gaussian prior
on the new candidate grid,
\begin{equation}
f_0\left(\bm{c}_{t_1^{(m)}}^j\right)
= \sum_{i=1}^{N_s}
  \gamma_{t_T^{(m-1)}}(i)\;
  f_\tau\!\left(\bm{c}_{t_1^{(m)}}^j,\;
                \bm{c}_{t_T^{(m-1)}}^i\right),
\label{eq:prior_carryover}
\end{equation}
followed by normalisation. In Equation~(\ref{eq:prior_carryover}), the prior at each candidate $\bm{c}^j_{t_1^{(m)}} \in \mathcal{S}_{t_1^{(m)}}$ is a sum over the candidates $\bm{c}^i_{t_T^{(m-1)}} \in \mathcal{S}_{t_T^{(m-1)}}$, weighted by $\gamma_{t_T^{(m-1)}}(i)$. In general, $\mathcal{S}_{t_1^{(m)}} \neq \mathcal{S}_{t_T^{(m-1)}}$, because the grid centre of Equation~(\ref{eq:grid_centre}) differs between $t_T^{(m-1)}$ and $t_1^{(m)}$. The transition factor $f_\tau(\bm{s}_{t_k}, \bm{s}_{t_{k-1}}) = p(\bm{s}_{t_k} \mid \bm{s}_{t_{k-1}})$ of Section~\ref{SubSec:FGChain} is defined for every pair of positions. Accordingly, Equation~(\ref{eq:prior_carryover}) requires $\gamma_{t_T^{(m-1)}}$ only on $\mathcal{S}_{t_T^{(m-1)}}$, where it is computed, and no interpolation between $\mathcal{S}_{t_T^{(m-1)}}$ and $\mathcal{S}_{t_1^{(m)}}$ is needed. Relative to $\gamma_{t_T^{(m-1)}}$, the prior of Equation~(\ref{eq:prior_carryover}) is broadened by a single transition step, whose standard deviation on the representative flight of Section~\ref{Sec:Validation} is 16~m per axis, approximately half a cell (see Section~\ref{SubSec:Trans}). Posterior mass that the transition step moves outside the region spanned by $\mathcal{S}_{t_1^{(m)}}$ is not represented on the new grid and is removed by the normalisation. At all $5 \times 10^3$ segment boundaries of each ensemble in Section~\ref{Sec:Performance}, 100\% of the terminal posterior mass lies on candidates $\bm{c}^i_{t_T^{(m-1)}}$ whose predicted position $\bm{c}^i_{t_T^{(m-1)}} + \bm{v}_{t_T^{(m-1)}}\,\Delta t$ [see Equation~(\ref{eq:motion_model})] falls within the region spanned by $\mathcal{S}_{t_1^{(m)}}$.

\subsection{Transition factor}
\label{SubSec:Trans}

The transition factor $f_\tau$ encodes the aircraft dynamics between
successive hidden states. Let $\Delta t = t_k - t_{k-1}$ and
$\bm{v}_{t_{k-1}} = [v^{\rm (N)}_{t_{k-1}}, v^{\rm (E)}_{t_{k-1}}]$ denote the sampling
interval and the INS
velocity estimate at time $t_{k-1}$, respectively. The motion model,
\begin{equation}
\bm{s}_{t_k}
= \bm{s}_{t_{k-1}}
  + \bm{v}_{t_{k-1}}\,\Delta t
  + \bm{\eta}_{t_k},
\label{eq:motion_model}
\end{equation}
assumes that the true displacement is the integrated INS velocity plus a
zero-mean Gaussian perturbation
$\bm{\eta}_{t_k}
\sim \mathcal{N}\!\bigl(\bm{0},\,\sigma_v^2\,\Delta t^2\,\bm{I}_2\bigr)$,
where $\sigma_v$ parameterises the velocity uncertainty per axis. The
transition factor,
\begin{equation}
f_\tau\left(\bm{c}_{t_k}^j,\,\bm{c}_{t_{k-1}}^i\right)
= \exp\!\left\{
  -\frac{\|\bm{d}_{ij} - \bm{v}_{t_{k-1}}\,\Delta t\|^2}
        {2\,\sigma_v^2\,\Delta t^2}
\right\},
\label{eq:trans_magnav}
\end{equation}
evaluates the displacement
$\bm{d}_{ij} = [d_{ij}^{\rm (N)},\,d_{ij}^{\rm (E)}]$ from candidate
$\bm{c}_{t_{k-1}}^i$ to candidate $\bm{c}_{t_k}^j$ against the predicted
displacement $\bm{v}_{t_{k-1}}\,\Delta t$,
with north and east displacement components given by
\begin{equation}
d_{ij}^{\rm (N)}
= (\phi_{t_k}^j - \phi_{t_{k-1}}^i)\,R_\oplus,
\qquad
d_{ij}^{\rm (E)}
= (\lambda_{t_k}^j - \lambda_{t_{k-1}}^i)\,R_\oplus\,
  \cos\phi_{t_{k-1}}^i,
\label{eq:displacement_NE}
\end{equation}
where $R_{\oplus}$ denotes the mean Earth radius. The differences $(\phi_{t_k}^j - \phi_{t_{k-1}}^i)$ and $(\lambda_{t_k}^j - \lambda_{t_{k-1}}^i)$ in Equation~(\ref{eq:displacement_NE}) are evaluated in radians.
\subsection{Measurement factor}
\label{SubSec:Meas}

The measurement factor $f_m$ quantifies the consistency between a candidate
position and the observed scalar magnetic anomaly. The observation model,
\begin{equation}
z_{t_k} = B_{\rm anom}(\bm{s}_{t_k}) + \epsilon_{t_k},
\label{eq:obs_model}
\end{equation}
assumes additive Gaussian noise
$\epsilon_{t_k} \sim \mathcal{N}(0,\,\sigma_{\rm mag}^2)$, where
$B_{\rm anom}(\bm{s}_{t_k})$ denotes the map value interpolated at the true position and
$\sigma_{\rm mag}$ absorbs contributions from sensor noise, map interpolation
error, and residual compensation artefacts. The measurement factor is
\begin{equation}
f_m\left(z_{t_k},\,\bm{c}_{t_k}^j\right)
= \exp\!\left\{
  -\frac{\left[z_{t_k} - B_{\rm anom}\left(\bm{c}_{t_k}^j\right)\right]^2}
        {2\,\sigma_{\rm mag}^2}
\right\}.
\label{eq:meas_magnav}
\end{equation}
Candidates whose map values closely match the observation receive factor values approaching unity; those predicting substantially different anomaly values are exponentially suppressed.

Extensions to vector magnetometry for improved aeromagnetic compensation  \citep{Lathrop_2024,Hager_2026} and navigation performance \citep{Shockley_2014, Canciani_2020, Li_2026} enter the factor graph formalism of Section \ref{Sec:FG} in at least two ways. With respect to the former, a vector magnetometer provides directional information about platform-generated interference, yielding a higher-fidelity scalar input $z_{t_{k}}$ without modification to Equation (\ref{eq:meas_magnav}). With respect to the latter, the scalar residual $[ z_{t_{k}} - B_{\rm anom}( \bm{c}^j_{t_{k}})]^2/\sigma_{\rm mag}^2$ in Equation (\ref{eq:meas_magnav}) is replaced by the Mahalanobis distance $[\bm{z}_{t_{k}} - \bm{R}_{t_{k}} \bm{B}_{\rm anom} ( \bm{c}^j_{t_{k}} ) ]^{\rm T} \, \bm{\Sigma}_{\rm mag}^{-1} \, [\bm{z}_{t_{k}} - \bm{R}_{t_{k}} \bm{B}_{\rm anom}( \bm{c}^j_{t_{k}} ) ]$, where $\bm{z}_{t_{k}} \in \mathbb{R}^{3}$ is the body-frame vector observation, $\bm{B}_{\rm anom} (\bm{c}^{j}_{t_{k}}) \in \mathbb{R}^{3}$ is a vector anomaly reference map, $\bm{R}_{t_{k}}$ is the map-to-body rotation matrix, and $\bm{\Sigma}_{\rm{mag}} \in \mathbb{R}^{3 \times 3}$ is the measurement noise covariance. The rotation matrix $\bm{R}_{t_{k}}$ couples the measurement factor to the INS attitude error states; see Section \ref{SubSec:SimEnv} for further details. Although a vector formulation fits readily into the factor graph framework of Section \ref{Sec:FG}, it requires a vector anomaly reference map as well as axis-resolved compensation. Accordingly, we retain the scalar formulation in this paper and defer vector extensions to future work.

\subsection{Model Idealisations}
\label{SubSec:Idealisations}

The factor graph formulation of Sections~\ref{SubSec:Discretisation}--\ref{SubSec:Meas} rests on several idealisations, three of which are discussed below. 

(i) The transition factor, Equation~(\ref{eq:trans_magnav}), assumes that the displacement noise
$\bm{\eta}_{t_k} \sim \mathcal{N}(\bm{0},\,
\sigma_v^2\,\Delta t^2\,\bm{I}_2)$ is isotropic, stationary, and independent between timesteps. In practice, INS velocity errors contain systematic biases that
grow with time and differ between the north and east axes. The per-step displacement residuals of the INS of Section~\ref{Sec:Validation} are strongly correlated, so the displacement error accumulated within a segment grows approximately $\propto T$ (see Section~\ref{SubSec:SelectingT}). The factor graph accounts for such departures in two places. First, $\sigma_v$ is an effective velocity uncertainty, estimated by maximising the Bayesian evidence, i.e.\ the marginal likelihood $p(z_{t_1:t_N} \mid \sigma_{\rm mag}, \sigma_v)$ of the measurements (see Section~\ref{SubSec:DataPrep}). Second, the upper bound on $T$ in Section~\ref{SubSec:SelectingT} allows for this growth. In Appendix~\ref{App:Sigma}, we formalise the estimation of $\sigma_v$ and $\sigma_{\rm mag}$. 

(ii) The measurement factor, Equation~(\ref{eq:meas_magnav}), assumes that
$\{\epsilon_{t_k}\}_{k=1}^{N} \sim \mathcal{N}(0, \sigma_{\rm mag}^2)$ are
independent and identically distributed residuals. In practice,
$\epsilon_{t_k}$ absorbs map interpolation error, uncompensated platform
interference, and time-varying external fields \citep{Hinze_2013}. Map
interpolation error is spatially correlated on the scale of
$\Delta_{\rm cell}$, and external-field residuals may exhibit correlation
times of minutes to hours. The parameter $\sigma_{\rm mag}$ is then an effective noise level, estimated with $\sigma_v$ in Section~\ref{SubSec:DataPrep}. Appendix~\ref{App:MapMismatch} measures the effect of a map error $\delta B$, the difference between the true anomaly field and the map, which is correlated along the track. Its results suggest that the median position error of the MMSE estimate is insensitive to $\delta B$ for $\delta B_{\rm rms} \ll \sigma_{\rm mag}$, but begins to increase for $\delta B_{\rm rms} \gtrsim \sigma_{\rm mag}$, where $\delta B_{\rm rms}$ is the root-mean-square value of $\delta B$ along the track.
 
(iii) The discretisation of the position domain onto the map grid
(Section~\ref{SubSec:Discretisation}) restricts the hidden state to cell centres spaced $\Delta_{\rm cell}$ apart. The MMSE estimate of Equation~(\ref{eq:mmse}) can lie between cell centres, while the marginal MAP estimate of Equation~(\ref{eq:map}) returns a cell centre from the same posterior, so the difference between their errors bounds the cost of quantisation. On the two ensembles of Section~\ref{Sec:Performance} this difference is 2.5~m and 3.1~m in per-trial median error (see Section~\ref{SubSec:Comparison} and Appendix~\ref{App:Viterbi}), small compared with the cell size. In regions where the anomaly field
varies on sub-grid scales, this structure is unresolvable. Conversely, where
$\|\nabla B_{\rm anom}\|$ is weak or spatially repetitive, the posterior remains diffuse
regardless of $T$, reflecting an identifiability limit of the
map-matching approach. Here we denote the
spatial gradient of the anomaly map $B_{\rm anom}(\phi, \lambda)$ by
$\nabla B_{\rm anom} = (\partial B_{\rm anom}/\partial\phi,\;\partial B_{\rm anom}/\partial\lambda)^{\rm{T}}$ along geodetic latitude $\phi$ and longitude $\lambda$ and $\nabla B_{\rm anom} = (\partial B_{\rm anom}/\partial N,\;\partial B_{\rm anom}/\partial E)^{\rm{T}}$ along local north and east coordinates where $\partial/\partial N = (1/R_{\oplus})\partial/\partial \phi$ and $\partial/\partial E = [1/(R_{\oplus} \cos\phi)]\partial/\partial \lambda$. Unless specified otherwise, we adopt the latter definition and express spatial gradients along local north and east components for the remainder of the paper. 

The results of Sections~\ref{Sec:Validation} and~\ref{Sec:Performance} are obtained under idealisations (i)--(iii), with their effects quantified in the sections named above.

\subsection{Segmented processing}
\label{SubSec:Segmentation}

A single factor graph over a whole flight of $N$ timesteps begins the backward sweep, viz.\ the recursion for $\beta_{t_k}$ of Equation~(\ref{eq:beta_rec}), only when the flight ends. The computational cost of this sweep is $O(N\,N_s^2)$. We segment the flight into $M$ non-overlapping segments of $T$ timesteps each, and let $\mathcal{T}^{(m)} = \{t_1^{(m)}, \ldots, t_T^{(m)}\}$ denote the timesteps of segment $m \in \{1, \ldots, M\}$. The backward sweep of segment $m$ starts at $t_T^{(m)}$, so its computational cost is $O(T\,N_s^2)$ and each estimate in $\mathcal{T}^{(m)}$ is delayed by at most $T\,\Delta t$.

Each segment is processed by the forward-backward algorithm of Section~\ref{SubSec:forward_backward}, and consecutive segments are connected only through the terminal posterior of segment $m-1$, via the prior carry-over of Equation~(\ref{eq:prior_carryover}) and the carried correction of Equation~(\ref{eq:carried_correction}). Within segment $m$, each marginal posterior $\gamma_t$ [where $t \in \mathcal{T}^{(m)}$] is conditioned on the prior carried from segment $m-1$ and on the measurements $z_{\mathcal{T}^{(m)}}$. The active-set pruning of Section~\ref{SubSec:Pruning} reduces the computational cost of each segment below $O(T\,N_s^2)$.

The non-overlapping segmentation adopted above is one possible strategy for bounding the computational cost of the backward sweep and the delay of each estimate. An alternative strategy based on a sliding segment window is presented in Appendix~\ref{App:Sliding} for the convenience of the reader. Its segments overlap, so every position estimate draws on later measurements within its segment, at a higher computational cost.

\subsection{Selecting $T$}\label{SubSec:SelectingT}

The segment length $T$ is a key input to the factor graph of
Section~\ref{Sec:FG}, whose value is subject to competing physical
constraints. In addition, only segment lengths that divide the full
subsampled flight exactly are admitted, so that the flight may be partitioned
into equal-length non-overlapping segments without discarding samples. The
value of $T$ is determined from a single nominal trajectory, together with
simulated INS and magnetic realisations, and is then held fixed across
all subsequent trials.

A lower bound on $T$ ensures that each segment accumulates sufficient
magnetic information to resolve position at the map-cell scale. For a scalar
anomaly field observed under additive Gaussian noise with variance
$\sigma_{\rm mag}^2$, the cumulative Fisher information over $T$ steps is
$J(T) \approx T\,\overline{\|\nabla B_{\rm anom}\|^2}\,/\,\sigma_{\rm mag}^2$, where
$\overline{\|\nabla B_{\rm anom}\|^2}$ denotes the mean squared gradient magnitude along
the flight path. Requiring the Cram\'{e}r--Rao position scale
$J(T)^{-1/2}$ to fall below the map grid spacing $\Delta_{\rm cell}$ yields
\begin{equation}
T_{\rm info}
= \left\lceil
    \frac{\sigma_{\rm mag}^2}
         {\overline{\|\nabla B_{\rm anom}\|^2}\;\Delta_{\rm cell}^2}
  \right\rceil.
\label{eq:T_info}
\end{equation}
The scalar quantity $J(T)$ treats the Fisher information as orientation-averaged, viz.\ $\overline{\|\nabla B_{\rm anom}\|^2}$, and hence can overestimate the information available in the least informative direction; see \cite{Kay_1993} and Appendix \ref{App:FIM} for a full Fisher information matrix formulation.  A second lower bound
$T_{\rm motion}
= \lceil n_{\rm cells}\,\Delta_{\rm cell}\,/\,(v\,\Delta t)\rceil$,
with $n_{\rm cells}=2.5$, requires the aircraft to traverse at least
$n_{\rm cells}$ map cells per segment so that spatial diversity in the
anomaly field is sampled. The effective lower bound is
$T_{\rm lower} = \max(T_{\rm info},\,T_{\rm motion})$.

An upper bound on $T$ keeps the true position inside the candidate grid. At each timestep $t_k$ of the nominal trajectory, we define the north--east displacement
residual
\begin{equation}
\bm{r}_k
= \begin{bmatrix}
(\phi_{t_{k+1}}^{\rm true} - \phi_{t_k}^{\rm true})\,R_\oplus - v_{t_k}^{\rm (N)}\Delta t \\[4pt]
(\lambda_{t_{k+1}}^{\rm true} - \lambda_{t_k}^{\rm true})\,R_\oplus\cos\phi_{t_k}^{\rm true}
 - v_{t_k}^{\rm (E)}\Delta t
\end{bmatrix}
\label{eq:transition_residual_vector}
\end{equation}
between the true displacement and that predicted by the INS velocity. Let $\bm{\Sigma}_{\rm step}$ denote the sample covariance of $\bm{r}_k$ over $3 \times 10^2$ realisations of the INS of Section~\ref{SubSec:Scenario} (Table~\ref{tab:ins_params}) on the nominal trajectory, where $\lambda_{\max}(\bm{\Sigma}_{\rm step})$ denotes its largest eigenvalue. In operation, $\bm{\Sigma}_{\rm step}$ can be estimated from a calibration flight with GNSS.

The per-step residuals are strongly correlated, with an autocorrelation of 0.93 at a lag of five timesteps and 0.75 at a lag of ten in each axis. For strongly correlated residuals, the covariance of the displacement error accumulated over $T$ steps is $\approx T^2\,\bm{\Sigma}_{\rm step}$. The 95\% confidence ellipse of a Gaussian error whose covariance is $T^2\,\bm{\Sigma}_{\rm step}$ lies within a circle with radius $r_{\rm grow}(T)$, viz.\ the residual growth radius
\begin{equation}
r_{\rm grow}\left(T\right)
= \left[
    \chi^2_{0.95}\,\lambda_{\max}(\bm{\Sigma}_{\rm step})
  \right]^{1/2} T,
\label{eq:growth_radius_model}
\end{equation}
where $\chi^2_{0.95}\approx 5.99$ is the 95\% quantile of the chi-squared distribution with two degrees of freedom. At $T = 40$, $r_{\rm grow}(T) = 763$~m, against an empirical 95th percentile of 721~m for the accumulated residual over the $3 \times 10^2$ realisations. Equation~(\ref{eq:growth_radius_model}) agrees with that percentile to within 7\% for $20 \leq T \leq 40$, the segment lengths considered here. A complete model of inertial error growth is given by \citet{Woodman_2007}.

By Equations~(\ref{eq:grid_centre}) and~(\ref{eq:carried_correction}), at each timestep of segment $m$ the offset of the true position from the grid centre is at most the displacement error accumulated over at most $T$ steps plus the error of the estimate at the preceding segment boundary, $e_{\rm b} = \|\hat{\bm{s}}^{\rm MMSE}_{t_T^{(m-1)}} - \bm{s}_{t_T^{(m-1)}}\|$. The first term is below $r_{\rm grow}(T)$ with probability at least 95\%, so the true position stays inside the grid when
\begin{equation}
r_{\rm grow}(T) + e_{\rm b} \leq n_{\rm half}\,\Delta_{\rm cell}.
\label{eq:containment}
\end{equation}

The segment length is an admissible $T$ satisfying
\begin{equation}
T \geq T_{\rm lower},
\qquad
r_{\rm grow}(T) + e_{\rm b} \leq n_{\rm half}\,\Delta_{\rm cell}.
\label{eq:T_selection_rule}
\end{equation}
For the values of $T$ that satisfy Equation~(\ref{eq:T_selection_rule}), a longer segment conditions each estimate on a larger number of later measurements, and a shorter segment reduces the delay $T\,\Delta t$. We adopt $T = 40$, for which the delay is at most $T\,\Delta t = 200$~s. At $T = 40$, $r_{\rm grow}(T)$ is about half of $n_{\rm half}\,\Delta_{\rm cell} = 1463$~m.

In each of the two ensembles of Section~\ref{Sec:Performance}, the true position stays inside the grid at every timestep. If the true position leaves the grid, the posterior remains normalised over the candidate set, so the estimate lies inside the grid and its error is at least the distance from the true position to the grid edge.

\subsection{Active-set pruning}
\label{SubSec:Pruning}

The forward and backward recursions in
Equations~(\ref{eq:alpha_rec})--(\ref{eq:beta_rec}) each require an
$O(N_s^2)$ summation per timestep. This cost can be reduced by excluding
candidates for which the measurement factor is negligible. At each timestep
$t_k$, the candidate
$j^* = \arg\max_j f_m(z_{t_k},\,\bm{c}_{t_k}^j)$ minimises the squared
residual $[z_{t_k} - B_{\rm anom}(\bm{c}_{t_k}^j)]^2$. The active set $\mathcal{A}_{t_k}$ retains only
those candidates whose measurement factor is at least $\alpha_{\rm prune} \in (0,1]$ times the maximum,
$f_m(z_{t_k},\,\bm{c}_{t_k}^{j^*})$. It is defined according to
\begin{equation}
\mathcal{A}_{t_k}
= \bigl\{\, j \in \{1,\ldots,N_s\} :
  \log f_m\left(z_{t_k},\,\bm{c}_{t_k}^j\right)
  \geq
  \log f_m\left(z_{t_k},\,\bm{c}_{t_k}^{j^*}\right) + \log\alpha_{\rm prune}
\,\bigr\}.
\label{eq:active_set}
\end{equation}
Substituting Equation~(\ref{eq:meas_magnav}) into Equation~(\ref{eq:active_set}), the condition retains all
candidates satisfying
$[z_{t_k} - B_{\rm anom}(\bm{c}_{t_k}^j)]^2
\leq [z_{t_k} - B_{\rm anom}(\bm{c}_{t_k}^{j^*})]^2
     - 2\sigma_{\rm mag}^2\log\alpha_{\rm prune}$,
i.e.\ those whose squared residual does not exceed that of $j^*$ by more than
$2\sigma_{\rm mag}^2|\log\alpha_{\rm prune}|$. The pruned forward recursion replaces the
sum over $\{1,\ldots,N_s\}$ in Equation~(\ref{eq:alpha_rec}) with a sum over
$\mathcal{A}_{t_{k-1}}$,
\begin{equation}
\alpha_{t_k}[j]
= f_m\left(z_{t_k},\,\bm{c}_{t_k}^j\right)
  \sum_{i \in \mathcal{A}_{t_{k-1}}}\left[
  \alpha_{t_{k-1}}[i]\,
  f_\tau\left(\bm{c}_{t_k}^j,\,\bm{c}_{t_{k-1}}^i\right)\right],
\quad j \in \mathcal{A}_{t_k},
\label{eq:alpha_pruned}
\end{equation}
and the backward recursion is modified analogously. If
$|\mathcal{A}| = \max_k |\mathcal{A}_{t_k}|$ denotes the maximum active-set
cardinality, the per-segment cost reduces from $O(T N_s^2)$ to
$O(T|\mathcal{A}|^2)$, with $|\mathcal{A}| \ll N_s$ in regions of strong spatial magnetic variation. The active set is rebuilt at every timestep, so a cell excluded at one timestep can return at the next. With $\alpha_{\rm prune} = 0.01$ (Table~\ref{tab:fg_params}), the cell nearest the true position is excluded from $\mathcal{A}_{t_k}$ at a median of 2.9\% and 1.7\% of the timesteps of a trial in the ensembles of Sections~\ref{SubSec:Noise} and~\ref{SubSec:Observability}, respectively. Appendix~\ref{App:ActiveSetPruning} measures the exclusion rate, the bias of $H_t$, the error, and the run time against $\alpha_{\rm prune}$.

\section{Validation with synthetic data}\label{Sec:Validation}

In this section, we present a representative magnetic anomaly navigation validation test using forward-backward inference on a chain-structured factor graph, conducted on synthetic INS, magnetometer, and trajectory data generated over a real aeromagnetic anomaly map. The simulation framework is described in Section~\ref{SubSec:SimEnv}. The scenario definition and its associated inputs and outputs are given in Sections~\ref{SubSec:Scenario} and~\ref{SubSec:DataPrep}, respectively. Some simplifying simulation idealisations and their associated effects are canvassed briefly in Section~\ref{SubSec:SimIdealisations}. A worked example of the forward-backward algorithm of Section~\ref{SubSec:forward_backward}, using the data generated synthetically in Section~\ref{SubSec:Scenario}, is presented in Section~\ref{SubSec:MagNavError}. We compare the Shannon entropy $H_t$ at time $t$ with Fisher-information diagnostics and navigation error in Section~\ref{SubSec:ShannonEntropy}. We summarise the forward-backward algorithm logic and pseudocode in Algorithm \ref{alg:fg_magnav} for the convenience of the reader. An overview of the Fisher information matrix \citep{Kay_1993} is presented in abridged form in Appendix \ref{App:FIM}.

\subsection{Simulation environment}
\label{SubSec:SimEnv}
 
All synthetic data employed in this paper are generated using the open-source Julia package
\texttt{MagNav.jl},\footnote{\href{https://github.com/MIT-AI-Accelerator/MagNav.jl}{https://github.com/MIT-AI-Accelerator/MagNav.jl}} developed by the Department of the Air Force--Massachusetts Institute of Technology Artificial
Intelligence Accelerator program for research in airborne MagNav.
The package has been validated against flight data collected in 2020 by Sanders Geophysics Ltd.\footnote{\href{https://www.sgl.com/}{https://www.sgl.com/}} near Ottawa,
Ontario, Canada and provides trajectory generation,
strapdown INS error propagation, scalar magnetometer simulation, aeromagnetic
compensation, as well as statistical signal processing algorithms. We summarise the components relevant
to this work below and we refer the reader to \cite{Gnadt_2022} for a detailed summary about the \texttt{MagNav.jl} Julia package and to \cite{Gnadt_2023} for instructions about accessing the Sanders Geophysics Ltd. 2020 flight data.
 
A ground-truth trajectory $\{\bm{s}_{t_k}\}_{k=1}^{N_{\rm sim}}$ is generated on a
real aeromagnetic anomaly map $B_{\rm anom}(\phi, \lambda)$ at prescribed airspeed $v$,
altitude $h$, flight duration $t_{\rm flight}$, and lateral wave structure.
The simulated trajectory is sampled at interval $\Delta t_{\rm sim}$.

The INS simulation follows the Pinson error model \citep{Titterdon_2004,Canciani_2016}, which propagates an 18-dimensional state vector $\bm{x}_{\rm INS}(t)$ comprising position error $\delta\bm{p}(t) \in \mathbb{R}^3$, velocity error $\delta\bm{v}(t) \in \mathbb{R}^3$, attitude (tilt) error $\bm{\epsilon}(t) \in \mathbb{R}^3$, accelerometer bias $\bm{b}_a(t) \in \mathbb{R}^3$, gyroscope bias $\bm{b}_g(t) \in \mathbb{R}^3$, barometric altimeter error $\delta h_b(t)$, vertical accelerometer error $\delta\hat{a}(t)$, and a world-frame time-correlated magnetic bias state $S(t)$.
 
The INS-measured specific force $\bm{f}_{\rm meas}(t)$ is related to the true specific force $\bm{f}_{\rm true}(t)$, viz.\ 
 \begin{equation}
 \bm{f}_{\rm meas}(t) = \bm{f}_{\rm true}(t) + \bm{b}_a(t) + \bm{\eta}_a(t).
 \label{Eq:forceMeas}
 \end{equation}
Similarly, the INS-measured angular rate $\bm{\omega}_{\rm meas}(t)$ is related to the true angular rate $\bm{\omega}_{\rm true}(t)$, viz.\ 
\begin{equation}
\bm{\omega}_{\rm meas}(t) = \bm{\omega}_{\rm true}(t) + \bm{b}_g(t) + \bm{\eta}_g(t).
\label{Eq:AngMeas}
\end{equation}
In Equations (\ref{Eq:forceMeas}) and (\ref{Eq:AngMeas}), we assume that $\bm{\eta}_a(t) \sim \mathcal{N}(\bm{0},\, \sigma_{\rm VRW}^2\, \bm{I}_3)$ and $\bm{\eta}_g(t) \sim \mathcal{N}(\bm{0},\, \sigma_{\rm ARW}^2\, \bm{I}_3)$, where  $\sigma_{\rm VRW}$ and $\sigma_{\rm ARW}$ are the velocity random walk (VRW) and angular random walk (ARW) spectral densities, respectively. 
 
We assume that the bias terms $\bm{b}_a(t)$ and $\bm{b}_g(t)$ in Equations~(\ref{Eq:forceMeas}) and (\ref{Eq:AngMeas}), together with the barometric altimeter error $\delta h_{b}(t)$ and the magnetic bias state $S(t)$, satisfy independent first-order Gauss–Markov (FOGM) processes,
 
\begin{equation}\dot{X}(t) = -\frac{X(t)}{\tau_X} + \sqrt{\frac{2\sigma_X^2}{\tau_X}}\, W(t),
\label{Eq:FOGMProcess}
\end{equation}
where $X(t) \in \{\bm{b}_a(t),\, \bm{b}_g(t),\, \delta h_b(t),\, S(t)\}$, $\tau_X \in \{\tau_{\rm a},\, \tau_{\rm g},\, \tau_{\rm b},\, \tau_{S}\}$ is the corresponding correlation time, $\sigma_X \in \{\sigma_{\rm a},\, \sigma_{\rm g},\, \sigma_{\rm b},\, \sigma_{S}\}$ is the steady-state standard deviation, and $W(t)$ is a unit white-noise driver. Initial conditions for position, velocity, and attitude are drawn from zero-mean Gaussian distributions with standard deviations $\sigma_{\rm pos,0}$, $\sigma_{\rm vel,0}$, and $\sigma_{\rm att,0}$, respectively. The INS output at each timestep is a corrupted position estimate $\hat{\bm{s}}_{t_k}^{\rm INS}$ and velocity estimate $\bm{v}_{t_k}$. The foregoing INS error model parameters are listed in Table~\ref{tab:ins_params}.
 
Scalar magnetometer measurements are generated synthetically, viz.\
\begin{equation}
z_{t_{k}} = B_{\rm anom}\left(\bm{s}_{t_k}\right) + S\left(t_k\right) + \eta_{t_{k}},
\end{equation}
where $\eta_{t_{k}} \sim \mathcal{N}(0, \sigma_{\rm meas}^2)$ is white measurement noise. The magnetic measurement parameters are listed in Table~\ref{tab:mag_params}.

\subsection{Scenario definition}
\label{SubSec:Scenario}
 
The simulation scenario is defined on the \texttt{Renfrew\_395} aeromagnetic anomaly map
available with \texttt{MagNav.jl}. A synthetic aircraft flies at constant
airspeed $v = 68\,\rm{m\,s^{-1}}$ and altitude $h = 1000\,\rm{m}$ for a
duration of $t_{\rm flight} = 1.2 \times 10^3\,\rm{s}$, following a sinusoidal lateral
profile with $N_{\rm waves} = 3$ cycles. The anomaly map is evaluated at the simulation altitude using the interpolation and altitude-continuation routines implemented in \texttt{MagNav.jl}. That is, the scalar map value $B_{\rm anom}(\mathbf{s}_{t_k})$ used to generate and evaluate measurements corresponds to the anomaly field interpolated at the aircraft's horizontal position and upward-continued to $h=1000\,{\rm m}$. The \texttt{Renfrew\_395} map has a mean map grid spacing of $\Delta_{\rm{cell}} \approx 29.3 \, \rm{m}$ ($\Delta_{\phi} \approx 2.65 \times 10^{-4}\,{}^\circ$ in latitude and $\Delta_{\lambda} \approx 3.73 \times 10^{-4}\,{}^\circ$ in longitude at a mean latitude of $\phi \approx 45.4{}^{\circ}$). The simulation sampling interval is
$\Delta t_{\rm sim} = 0.1\,\rm{s}$, yielding
$N_{\rm sim} = t_{\rm flight}/\Delta t_{\rm sim} = 1.2 \times 10^4$ samples. The
INS and magnetometer data are subsampled to $\Delta t_{\rm FG} = 5.0\,\rm{s}$ for
the factor graph, yielding $N_{\rm FG} = t_{\rm flight}/\Delta t_{\rm FG} = 240$ timesteps. The factor graph sampling interval $\Delta t_{\rm FG}$ and number of timesteps $N_{\rm FG}$ coincide with $\Delta t$ and $N$ in Section~\ref{Sec:Magnav}.  The INS error model parameters are listed in Table~\ref{tab:ins_params} and
the magnetic measurement parameters in Table~\ref{tab:mag_params}. A single
realisation of the INS and magnetic noise processes is drawn with a fixed
random seed; ensemble statistics are reported in
Section~\ref{Sec:Performance}.

\begin{figure}
\centering
\includegraphics[width=\linewidth]{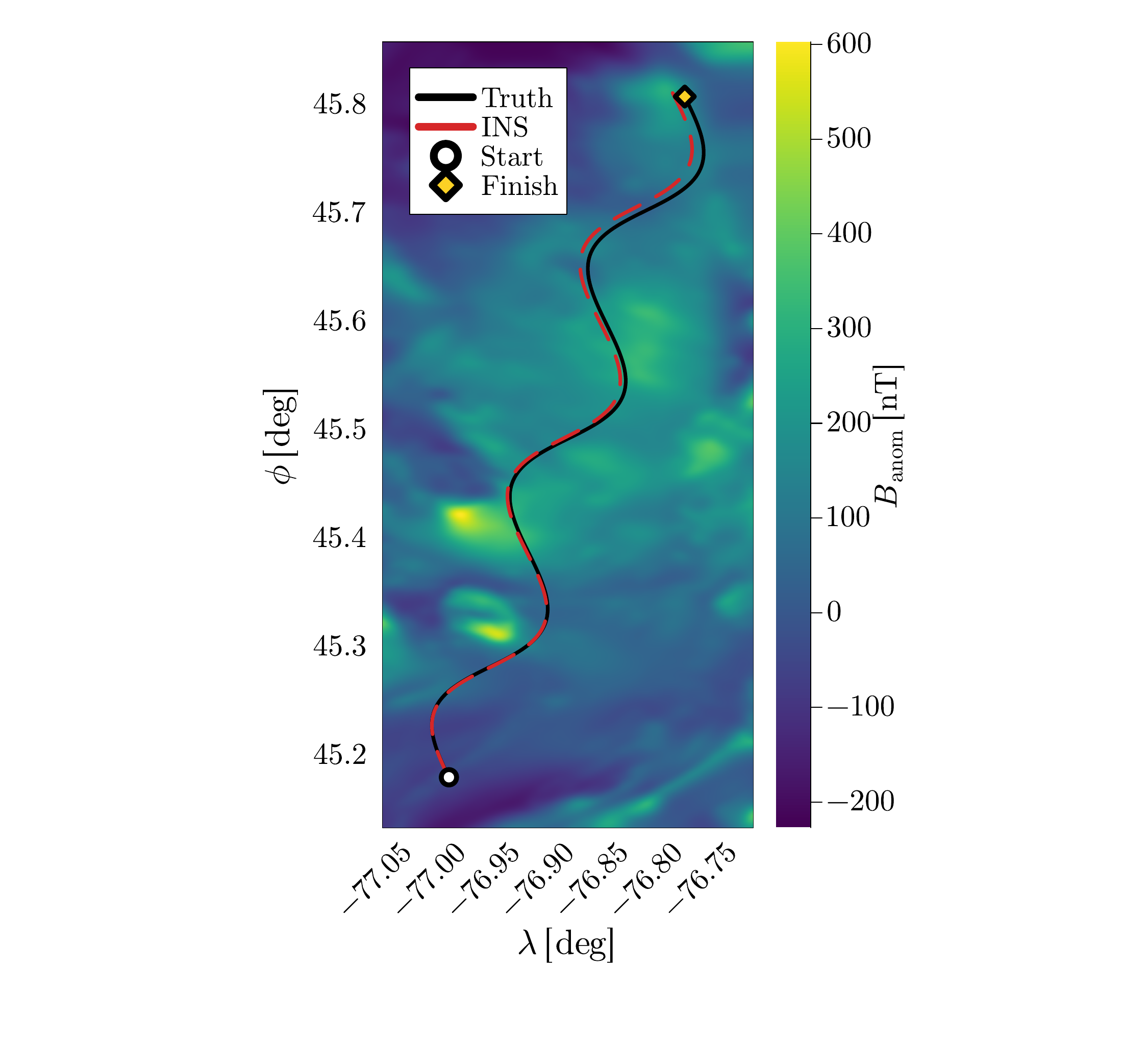}
\caption{The \texttt{Renfrew\_395} aeromagnetic anomaly map $B_{\rm anom}(\phi, \lambda)$ (units: nT), overplotted with the ground-truth trajectory $\bm{s}_{t_k}$ (black, solid curve) and the
  INS position estimate $\hat{\bm{s}}_{t_k}^{\rm INS}$ (red, dashed curve). The start and
  finish positions are marked with a white circle and yellow diamond, respectively. Geodetic
  longitude $\lambda$ (units: degrees) is reported on the horizontal axis and geodetic
  latitude $\phi$ (units: degrees) on the vertical axis. The colour bar spans the anomaly field at the flight altitude $h = 1000$~m over the plotted region, from its smallest to its largest map value}
\label{Fig:MagNavMap}
\end{figure}

In Figure~\ref{Fig:MagNavMap}, we plot the \texttt{Renfrew\_395} aeromagnetic anomaly map $B_{\rm anom}(\phi,
  \lambda)$ (units: $\rm{nT}$), overplotted with the ground-truth trajectory $\bm{s}_{t_k}$ (black,
  solid curve) and the INS position estimate $\hat{\bm{s}}_{t_{k}}^{\rm INS}$ (red, dashed curve). The start and finish positions are marked with a white circle and yellow diamond, respectively. The horizontal and vertical axes denote geodetic longitude $\lambda$ (units: degrees) and latitude $\phi$ (units: degrees), respectively. The colour bar spans the anomaly field at the flight altitude over the plotted region, $-2 \lesssim B_{\rm anom}(\phi,\lambda)/(100 \,  \rm{nT}) \lesssim 6$, with limits set by the smallest and largest map values in that region. This range is fixed by the local geology and by the attenuation of short wavelengths in the upward continuation from the survey altitude of 395~m to $h$.

  Figure~\ref{Fig:MagNavMap} has three key features. First, the ground-truth trajectory traverses
  approximately $0.63^\circ$ in latitude $(\sim 69.8 \, \rm{km})$ from south to north, with a sinusoidal
  lateral profile of $N_{\rm waves} = 3$ cycles. Second, the anomaly field exhibits coherent spatial
  structure on scales of $\sim 0.05^\circ$--$0.1^\circ$. Localised high-gradient features, e.g.\ near $\phi
  \approx 45.4^\circ$, are interspersed with smoother regions, e.g.\ near $45.6^{\circ} \lesssim \phi \lesssim 45.7^{\circ}$. Third, the INS trajectory $\hat{\bm{s}}_{t_{k}}^{\rm INS}$ diverges progressively from the truth, with the separation growing to $\sim 1.07$ km by the end of the flight.

\begin{table}[t]
\centering
\caption{INS error model parameters for the representative realisation.}
\label{tab:ins_params}
\begin{tabular}{llll}
\toprule
{\bfseries Parameter} & {\bfseries Symbol} & {\bfseries Value} & {\bfseries Units} \\
\midrule
Initial position uncertainty
  & $\sigma_{\rm pos,0}$ & 100 & m \\
Initial velocity uncertainty
  & $\sigma_{\rm vel,0}$ & 0.01 & m\,s$^{-1}$ \\
Initial attitude uncertainty
  & $\sigma_{\rm att,0}$ & 0.01 & deg \\
Velocity random walk
  & $\sigma_{\rm VRW}$ & $2.38 \times 10^{-4}$ & m\,s$^{-3/2}$ \\
Angular random walk
  & $\sigma_{\rm ARW}$ & $5.81 \times 10^{-7}$ & rad\,s$^{-1/2}$ \\
Accelerometer bias
  & $\sigma_{\rm a}$ & $2.45 \times 10^{-4}$ & m\,s$^{-2}$ \\
Gyroscope bias
  & $\sigma_{\rm g}$ & $7.27 \times 10^{-9}$ & rad\,s$^{-1}$ \\
Barometric altimeter noise
  & $\sigma_{\rm b}$ & 1.0 & m \\
Bias correlation times
  & $\tau_{\rm a},\,\tau_{\rm g},\,\tau_{\rm b}$ & 3600 & s \\
\bottomrule
\end{tabular}
\end{table}
 
\begin{table}[t]
\centering
\caption{Magnetic measurement model parameters.}
\label{tab:mag_params}
\begin{tabular}{llll}
\toprule
{\bfseries Parameter} & {\bfseries Symbol} & {\bfseries Value} & {\bfseries Units} \\
\midrule
White measurement noise variance
  & $\sigma_{\rm meas}^2$ & 1.0 & nT$^2$ \\
FOGM amplitude
  & $\sigma_{S}$ & 1.0 & nT \\
FOGM correlation time
  & $\tau_{S}$ & 600 & s \\
\bottomrule
\end{tabular}
\end{table}

\subsection{Data preparation}
\label{SubSec:DataPrep}
 
Before entering the factor graph, the simulation outputs undergo two preparation steps, which yield the algorithm inputs and the noise parameters of Section~\ref{Sec:Magnav}: (i) the magnetic measurements are debiased against the map along a reference path; and (ii) the noise parameters $\sigma_{\rm mag}$ and $\sigma_v$ are estimated by maximising the Bayesian evidence. The foregoing preparation steps do not employ the true trajectory $\{\bm{s}_{t_k}\}_{k=1}^{N_{\rm FG}}$.

The FOGM-biased magnetic measurements are subsampled to the factor graph cadence $\Delta t_{\rm FG}$, yielding $\{z_{t_k}^{\rm biased}\}_{k=1}^{N_{\rm FG}}$, and debiased against the map values along a reference path $\{\bar{\bm{s}}_{t_k}\}_{k=1}^{N_{\rm FG}}$, viz.\
\begin{equation}
z_{t_k} = z_{t_k}^{\rm biased} - \frac{1}{N_{\rm FG}} \sum_{l=1}^{N_{\rm FG}} \left[z_{t_l}^{\rm biased} - B_{\rm anom}\left(\bar{\bm{s}}_{t_l}\right)\right].
\label{eq:debias}
\end{equation}
Equation~(\ref{eq:debias}) is applied in successive passes, because the map values along a path displaced from the true trajectory differ from $B_{\rm anom}(\bm{s}_{t_k})$ by an amount set by the displacement and the local gradient $\nabla B_{\rm anom}$. On the first pass, the reference path is the INS position estimate, i.e.\ $\bar{\bm{s}}_{t_k} = \hat{\bm{s}}^{\rm INS}_{t_k}$, the only path available before the factor graph is run. On each subsequent pass, the reference path is the MMSE estimate $\hat{\bm{s}}^{\rm MMSE}_{t_k}$ of Equation~(\ref{eq:mmse}), which the factor graph returns from the measurements debiased on the preceding pass. The passes, the noise parameters used within each pass, and the stopping rule are specified in Appendix~\ref{App:Sigma}.

The noise parameters are estimated by maximising the Bayesian evidence, i.e.\ the marginal likelihood of the debiased measurements $z_{t_1:t_{N_{\rm FG}}}$, viz.\
\begin{equation}
(\hat{\sigma}_{\rm mag}, \hat{\sigma}_v) = \arg\max_{\sigma_{\rm mag},\,\sigma_v} p(z_{t_1:t_{N_{\rm FG}}} \mid \sigma_{\rm mag}, \sigma_v).
\label{eq:evidence_max}
\end{equation}
Let $\tilde{\alpha}_{t_k}$ denote the forward variable at $t_k$ before the renormalisation of Section~\ref{SubSec:forward_backward}, and let $\zeta_{t_k} = \sum_{j=1}^{N_s} \tilde{\alpha}_{t_k}[j]$ denote its normalising constant. The forward recursion of Equations~(\ref{eq:alpha_init}) and~(\ref{eq:alpha_rec}) yields the evidence as the product of the normalising constants \citep[Section~V.A]{Rabiner_1989}, viz.\
\begin{equation}
p(z_{t_1:t_{N_{\rm FG}}} \mid \sigma_{\rm mag}, \sigma_v) = \prod_{k=1}^{N_{\rm FG}} \zeta_{t_k}.
\label{eq:evidence_product}
\end{equation}
In Equation~(\ref{eq:evidence_product}), the recursion runs over the $M$ segments of the flight, connected by the prior carry-over of Equation~(\ref{eq:prior_carryover}), and the factors $f_m$ and $f_\tau$ of Equations~(\ref{eq:meas_magnav}) and~(\ref{eq:trans_magnav}) are multiplied by their Gaussian normalising constants, $(2\pi\sigma_{\rm mag}^2)^{-1/2}$ and $(2\pi\sigma_v^2\,\Delta t^2)^{-1}$, respectively.\footnote{Without the normalising constants, every factor in Equations~(\ref{eq:trans_magnav}) and~(\ref{eq:meas_magnav}) increases towards unity as $\sigma_{\rm mag}$ and $\sigma_v$ increase, so the right-hand side of Equation~(\ref{eq:evidence_product}) increases monotonically with each parameter and Equation~(\ref{eq:evidence_max}) has no finite maximiser.} The maximisation in Equation~(\ref{eq:evidence_max}) is performed on a grid of $(\sigma_{\rm mag}, \sigma_v)$ values, specified in Appendix~\ref{App:Sigma}. For the representative test case, it yields $\hat{\sigma}_{\rm mag} = 0.82$~nT and $\hat{\sigma}_v = 3.26$~m\,s$^{-1}$ (Table~\ref{tab:fg_params}). These are effective parameters of the factor graph (Section~\ref{SubSec:Idealisations}), so $\hat{\sigma}_{\rm mag}$ does not equal the white measurement noise $\sigma_{\rm meas}$ of Table~\ref{tab:mag_params} in general.
\begin{table}[t]
\centering
\caption{Factor graph parameters for the representative test case.}
\label{tab:fg_params}
\begin{tabular}{llll}
\toprule
{\bfseries Parameter} & {\bfseries Symbol} & {\bfseries Value} & {\bfseries Units} \\
\midrule
Measurement noise (evidence maximum)
  & $\sigma_{\rm mag}$ & $0.82$ & nT \\
Velocity uncertainty (evidence maximum)
  & $\sigma_v$ & $3.26$ & m\,s$^{-1}$ \\
Grid dimension
  & $n$ & $101$ & -- \\
Map grid spacing
  & $\Delta_{\rm cell}$ & $29.3$ & m \\
Segment length (derived)
  & $T$ & $40$ & -- \\
Sampling interval
  & $\Delta t_{\rm FG}$ & 5.0 & s \\
Pruning threshold
  & $\alpha_{\rm prune}$ & 0.01 & -- \\
\bottomrule
\end{tabular}
\end{table}

\subsection{Simulation idealisations}
\label{SubSec:SimIdealisations}

The validation scenario of Sections~\ref{SubSec:Scenario}--\ref{SubSec:DataPrep} rests on several idealisations associated with the simulation environment and parameter estimation, three of which are discussed below.

(i) The measurements are generated at a constant, known altitude $h = 1000\,\rm{m}$ from the same aeromagnetic anomaly map (\texttt{Renfrew\_395}), interpolation, and upward continuation as the measurement factor of Equation~(\ref{eq:meas_magnav}). The position domain is discretised in the horizontal plane $(\phi, \lambda)$ only, and the map error $\delta B$ of Section~\ref{SubSec:Idealisations} is zero. In practice, $\delta B$ arises from survey and compilation errors and from an error $\Delta h$ in the assumed altitude, viz.\ $\delta B \approx (\partial B_{\rm anom}/\partial h)\,\Delta h$. Altitude uncertainty from barometric altimeter errors or terrain-following flight profiles requires an independent altitude estimate or extension of the candidate grid to $n_z$ altitude levels. Such extensions increase $N_s$ from $n^2$ to $n_z n^2$ and the transition cost by a factor of up to $n_z^2$ (see Section~\ref{Sec:Conclusion}). Appendix~\ref{App:MapMismatch} quantifies the effect of $\delta B$, with measurements generated from a second survey of the same area (\texttt{Renfrew\_555}) and with the map continued to $\Delta h = 50$~m and 100~m above the flight altitude.

(ii) The bias removed by Equation~(\ref{eq:debias}) and the noise parameters $\sigma_{\rm mag}$ and $\sigma_v$ are estimated once per flight from the data of the whole flight (Section~\ref{SubSec:DataPrep}), and are held constant over the flight. The delay of at most $T\,\Delta t$ of Section~\ref{SubSec:Segmentation} holds at these fixed parameters. An operational system would estimate the bias and the noise parameters from the segments already flown (see Section~\ref{Sec:Conclusion}). Appendix~\ref{App:Sigma} examines the sensitivity of the position error to $\sigma_{\rm mag}$ and $\sigma_v$.

(iii) The aircraft airspeed $v = 68\,\rm{m\,s^{-1}}$ and lateral wave structure ($N_{\rm waves} = 3$ cycles) are held fixed across all trials. Different platform kinematics would alter the spatial sampling of the anomaly field and hence the segment-level observability and navigation performance. The ensemble of Section~\ref{SubSec:Observability} varies the heading and start location of the flight at fixed airspeed and lateral profile, and reports the resulting variation of the median position error of the MMSE estimate.

As with the model idealisations of Section~\ref{SubSec:Idealisations}, the results of Sections~\ref{Sec:Validation} and~\ref{Sec:Performance} are obtained under idealisations (i)--(iii), with their effects quantified in the appendices and sections named above.

\begin{algorithm}[H]
\caption{Segmented factor-graph inference for MagNav}
\label{alg:fg_magnav}
\begin{algorithmic}[1]
\Input $n_{\rm half}$, $T$, $\alpha_{\rm prune}$,
       $\hat{\sigma}_{\rm mag}$ and $\hat{\sigma}_v$ (Equation~\ref{eq:evidence_max}),
       $\{\hat{\bm{s}}^{\rm INS}_{t_k}\}_{k=1}^{N_{\rm FG}}$,
       $\{\bm{v}_{t_k}\}_{k=1}^{N_{\rm FG}}$,
       $\{z_{t_k}\}_{k=1}^{N_{\rm FG}}$ (Equation~\ref{eq:debias}), $B_{\rm anom}(\phi, \lambda)$
\State Set $n = 2n_{\rm half}+1$ and $M = N_{\rm FG}/T$
\State Partition the flight into $M$ non-overlapping segments of length $T$
\State Set the carried correction $\Delta\hat{\bm{s}}^{(0)} = \bm{0}$
\For{$m = 1,\ldots,M$}
    \For{$t \in \mathcal{T}^{(m)}$}
        \State \algparbox{Centre an $n \times n$ candidate grid on the map node
               $\mathcal{Q}[\hat{\bm{s}}^{\rm INS}_{t} + \Delta\hat{\bm{s}}^{(m-1)}]$
               (Equation~\ref{eq:grid_centre})}
        \State Read the map value $B_{\rm anom}(\bm{c}_{t}^j)$ at each candidate
    \EndFor
    \If{$m=1$}
        \State Set the prior $f_0(\bm{c}_{t_1^{(1)}}^j) = 1/N_s$ for all $j$
    \Else
        \State \algparbox{Propagate the terminal posterior of segment $m-1$
               through one transition step onto the first grid of segment $m$
               and normalise (Equation~\ref{eq:prior_carryover})}
    \EndIf
    \For{$t \in \mathcal{T}^{(m)}$}
        \State Form the active set $\mathcal{A}_{t}$
               (Equation~\ref{eq:active_set})
    \EndFor
    \State Run the pruned forward recursion over $\mathcal{T}^{(m)}$
           (Equation~\ref{eq:alpha_pruned})
    \State Run the pruned backward recursion over $\mathcal{T}^{(m)}$
    \State \algparbox{Combine the forward and backward variables to form the
           marginal posteriors $\{\gamma_t\}_{t \in \mathcal{T}^{(m)}}$
           (Equation~\ref{eq:gamma})}
    \State \algparbox{Extract $\hat{\bm{s}}_{t}^{\rm MMSE}$
           (Equation~\ref{eq:mmse}) and $H_t$
           (Equation~\ref{eq:entropy}) at each timestep}
    \State \algparbox{Update the carried correction $\Delta\hat{\bm{s}}^{(m)}$
           from the MMSE estimate at $t_T^{(m)}$
           (Equation~\ref{eq:carried_correction})}
\EndFor
\Output $\{\hat{\bm{s}}_{t_k}^{\rm MMSE}\}_{k=1}^{N_{\rm FG}}$,
        $\{H_{t_k}\}_{k=1}^{N_{\rm FG}}$
\end{algorithmic}
\end{algorithm}

\subsection{Airborne magnetic anomaly navigation error}\label{SubSec:MagNavError}

\begin{figure}
    \centering
    \includegraphics[width=1.0\textwidth]{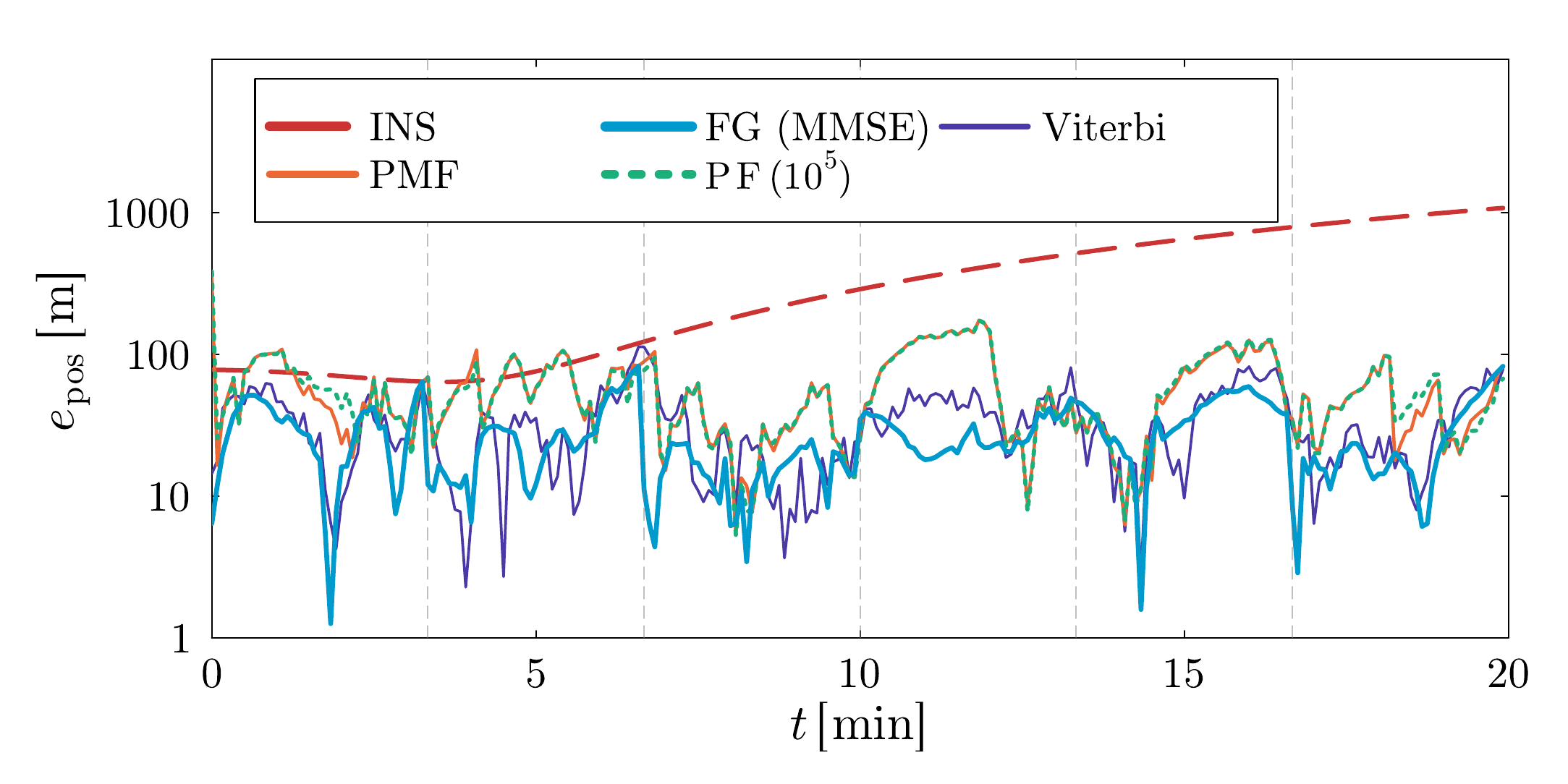}
    \caption{Position error $e_{\rm pos}(t)$ (units: m) versus time $t$ (units: min) for the
  representative test case of Section~\ref{SubSec:Scenario}. The error of the INS position estimate is
  plotted as a red, dashed curve and that of the forward-backward MMSE estimate
  (Equation~\ref{eq:mmse}) as a cyan, solid curve. The errors of the forward-only PMF (orange, solid curve), the bootstrap particle filter with $10^5$ particles (green, dotted curve), and the Viterbi path (purple, solid curve) of Appendix~\ref{App:Viterbi} are overplotted for comparison. The $M = N_{\rm FG}/T = 6$ segment boundaries
  (Section~\ref{SubSec:Segmentation}) are overplotted as grey, dashed, vertical lines. The vertical axis is plotted on a logarithmic scale and the horizontal axis on a linear scale. The factor graph parameters adopted in this section are
  listed in Table~\ref{tab:fg_params}}
    \label{Fig:ErrorTracking}
\end{figure}

The forward-backward algorithm of Section~\ref{SubSec:forward_backward} yields the marginal
  posterior $\gamma_{t_k}(j)$ of candidate $\boldsymbol{c}^j_{t_k}$ at time $t_k$ via the sum-product
  recursion of Equations (\ref{eq:gamma})--(\ref{eq:beta_rec}), the MMSE estimate of which is
  calculated according to Equation~(\ref{eq:mmse}). In Figure~\ref{Fig:ErrorTracking}, we plot the
  magnetic anomaly airborne position error $e_{\rm pos}(t) = \| \hat{\bm{s}}_t - \bm{s}_t \|$ (units: m) versus time $t$ (units: min) of the unaided INS (red, dashed curve) and the forward-backward MMSE estimate (cyan, solid curve). By way of additional comparison, we also plot the position error $e_{\rm pos}(t)$ of a forward-only PMF (orange, solid curve), the bootstrap particle filter with $10^5$ particles (green, dotted curve), and the Viterbi path (purple, solid curve). We summarise some details associated with the three foregoing estimators in Appendix~\ref{App:Viterbi}. Overplotted as grey, dashed, vertical lines are the $M = N_{\rm FG}/T = 6$ segment boundaries
  described in Section~\ref{SubSec:Segmentation}; see Table~\ref{tab:fg_params} for details about the
  factor graph values adopted in this section. The vertical axis is plotted on a logarithmic scale and the horizontal axis on a linear scale.

  Figure~\ref{Fig:ErrorTracking} has four key features. First, the INS position error starts at $77.7 \, \rm{m}$, a draw of the initial position error with standard deviation $\sigma_{\rm pos,0}$ (Table~\ref{tab:ins_params}). It reaches its minimum of $63.8 \, \rm{m}$ at $t = 3.5 \, \rm{min}$ and then grows unbounded to $\approx 1074 \, \rm{m}$ after $t_{\rm flight} = 1200 \, \rm{s}$ of flight. This growth is consistent with the expected $\propto t^2$ (accelerometer bias) and $\propto t^3$ (gyroscope bias) contributions to inertial drift \citep{Woodman_2007}. Second, the MMSE position error remains bounded in this representative realisation, with a median of $23.5 \, \rm{m}$ over the flight, to be compared with $284.9 \, \rm{m}$ for the INS. Third, the largest MMSE error, $82.7 \, \rm{m}$, occurs at $t = 6.6 \, \rm{min}$, the last timestep $t_T^{(2)}$ of segment 2. At the last timestep of each segment, $\beta_{t_T} = 1$ [Equation~(\ref{eq:beta_init})], so $\gamma_{t_T}$ equals the normalised forward variable $\alpha_{t_T}$, i.e.\ the filtering density of the PMF, and the MMSE and PMF errors coincide. At each of the ten timesteps with $e_{\rm pos}(t) > 2\,\Delta_{\rm cell} = 58.5 \, \rm{m}$, including $t_T^{(2)}$, between 90\% and 97\% of the posterior mass lies within $3\,\Delta_{\rm cell}$ of the MMSE estimate, i.e.\ the posterior is concentrated on a single mode displaced from the true position. At the next timestep, $t_1^{(3)}$, the MMSE estimate is conditioned on the measurements $z_{\mathcal{T}^{(3)}}$ of segment 3, and $e_{\rm pos}(t)$ falls to $11.1 \, \rm{m}$. Fourth, the PMF and the particle filter track each other closely. The PMF approximates the filtering density $p(\bm{s}_{t_k} \mid z_{t_1:t_k})$ on a grid, and the particle filter approximates the same density by samples. The median error is $48.6 \, \rm{m}$ for the PMF and $50.9 \, \rm{m}$ for the particle filter. The Viterbi path reaches a median error of $30.4 \, \rm{m}$.

  For the representative test case above, the active set has a median cardinality $|\mathcal{A}_{t_k}| = 674$ out of $N_s = n^2 = 10\,201$ candidates, and its 95th percentile is 2190. The pruned recursions evaluate $|\mathcal{A}_{t_{k-1}}|\,|\mathcal{A}_{t_k}|$ transitions per timestep in place of $N_s^2 \approx 1.04 \times 10^8$. The median number of transitions per timestep is $4.5 \times 10^5$, a reduction by a factor of $\approx 230$.

\subsection{Shannon entropy}\label{SubSec:ShannonEntropy}

The Shannon entropy $H_t$ of the marginal posterior [Equation~(\ref{eq:entropy})] quantifies the uncertainty in the position estimate at each timestep as a single scalar, with $H_t = 0$ indicating a degenerate posterior concentrated on one candidate and $H_t = \ln N_s$ indicating a uniform distribution over
  all candidates. The Shannon entropy $H_t$ is computed from $\gamma_t$ and hence it is available operationally. For the representative test case above, we find that $1.14 \leq H_t/(1\,{\rm nat}) \leq 3.14$, i.e.\ $0.12 \leq \bar{H}_t \leq 0.34$ [Equation~(\ref{eq:entropy_norm})], with a mean of 2.15 nats ($\bar{H}_t = 0.23$), for $\ln N_s = 9.23$.

  For the scalar Gaussian measurement model of Equation~(\ref{eq:obs_model}), the Fisher information matrix at
  position $\bm{s}_t$ is $J_t = \sigma_{\rm mag}^{-2}  \nabla B_{\rm anom}  (\nabla B_{\rm anom})^{\rm T}$ (see Appendix
  \ref{App:FIM}), with trace $\operatorname{tr}(J_t) = \|\nabla B_{\rm anom}\|^2 / \sigma_{\rm mag}^2$. Throughout this section and Section~\ref{Sec:Performance}, the gradient $\nabla B_{\rm anom}$ is evaluated in local north-east coordinates (units: nT $\rm{km^{-1}}$) so that the Fisher information and its eigenvalues have units of $\rm{km}^{-2}$. For this validation diagnostic, $\nabla B_{\rm anom}$ is evaluated along the known truth trajectory. Because $J_t$ is rank-1, its minimum eigenvalue is identically zero, and a single scalar observation cannot resolve position
  along the direction orthogonal to $\nabla B_{\rm anom}$. To characterise the information available from
  multiple measurements, we define the per-segment Fisher information $J^{(m)} = \sum_{t \in \mathcal{T}^{(m)}}
  J_t$ over the $T = 40$ timesteps of segment $m$. Provided the measurements
  within $\mathcal{T}^{(m)}$ are independent (consistent with the noise model of Section~\ref{SubSec:Meas}) and the
  aircraft trajectory samples sufficiently distinct gradient directions, $J^{(m)}$ is full-rank and its
  minimum eigenvalue $\lambda_{\min}[J^{(m)}]$ identifies the least informative spatial direction over the
  segment.

\begin{figure}
  \centering
  \includegraphics[width=1\textwidth]{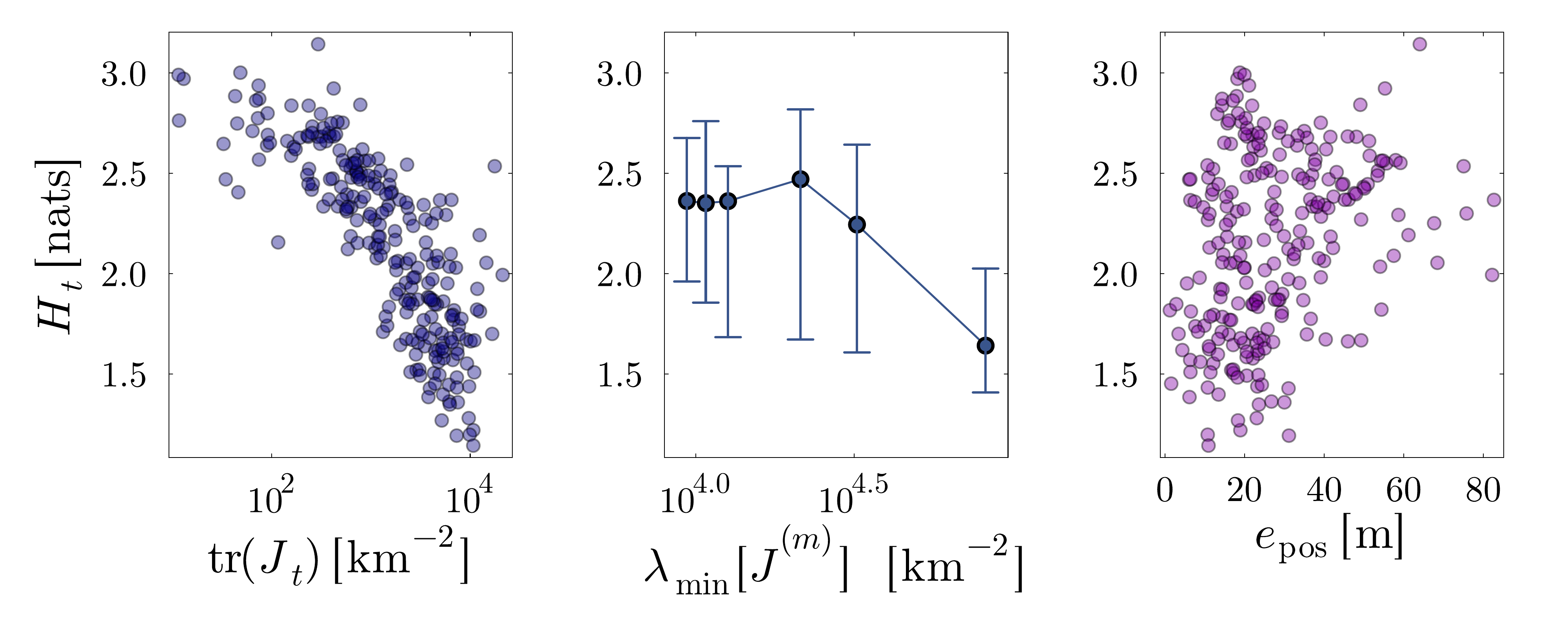}
  \caption{Shannon posterior entropy $H_t$ [Equation~(\ref{eq:entropy}); units: nats] for the representative test case of
  Section~\ref{SubSec:Scenario}. Left panel:
  $H_t$ versus the per-timestep Fisher trace $\operatorname{tr}(J_t)$ (units: km$^{-2}$) on
  a logarithmic horizontal axis, with one point per timestep ($N_{\rm FG} = 240$). Middle panel:
  segment-level median entropy $H^{(m)}_{\rm med} = \operatorname{median}_{t \in \mathcal{T}^{(m)}} H_t$ versus the segment-level
  minimum eigenvalue of the accumulated Fisher information $\lambda_{\min}[J^{(m)}]$ (units:
  km$^{-2}$) on a logarithmic horizontal axis, with one point per segment ($M = 6$). The
  vertical bars denote the 16th to 84th percentile  of $H_t$ within each segment.
  Right panel: $H_t$ versus position error $e_{\rm pos}(t)$ (units: m) on a linear
  horizontal axis, with one point per timestep ($N_{\rm FG} = 240$)}
  \label{Fig:EntropyPlots}
  \end{figure}

  In Figure~\ref{Fig:EntropyPlots}, we plot the Shannon entropy $H_t$ (units: nats) against three diagnostic
  quantities, using the 240 per-timestep samples and 6 segment summaries from the representative test
  case. In the left panel, $H_t$ is plotted against the per-timestep Fisher trace $\operatorname{tr}
  (J_t)$ (km$^{-2}$) on a logarithmic horizontal axis; both quantities are evaluated at each of the
  240 timesteps. In the middle panel, the segment-level median entropy, $H^{(m)}_{\rm med} = \operatorname{median}_{t \in \mathcal{T}^{(m)}} H_t$, is plotted against the segment-level minimum eigenvalue $\lambda_{\min}[J^{(m)}]$ (km$^{-2}$) on a
  logarithmic horizontal axis, with one point per segment ($M = 6$) and vertical bars spanning the
  16th to 84th percentile range of $H_t$ within each segment. In the right panel, $H_t$ is plotted
  against the position error $e_{\rm pos}(t)$ (units: m) on a linear horizontal axis. All three panels share a common vertical axis.

  Figure~\ref{Fig:EntropyPlots} has three key features. First, the left panel reveals a negative correlation between
  $H_t$ and $\log_{10} \operatorname{tr}(J_t)$, with $\rho\big[H_t, \log_{10}\operatorname{tr}(J_t)\big] = -0.80$, where $\rho(x, y)$ denotes the Pearson correlation coefficient between $x$ and $y$.
  The per-timestep Fisher trace spans approximately $11.5 \, \rm{km^{-2}}$ to $2.12 \times 10^4$ km$^{-2}$, with a
  median of $1.68 \times 10^3$ km$^{-2}$. Physically, timesteps at which the aircraft traverses regions of steep
  magnetic gradient [large $\operatorname{tr}(J_t)$] yield more informative measurement factors, which
  is consistent with the expectation that stronger measurement factors concentrate the marginal
  posterior. In smoother regions, the measurement factor is less discriminating and the posterior
  tends to be more diffuse. Second, the middle panel shows a generally decreasing trend of segment-
  level median entropy with increasing $\lambda_{\min}[J^{(m)}]$, with $\rho\big\{H^{(m)}_{\rm med}, \log_{10}\lambda_{\min}[J^{(m)}]\big\} = -0.84$ across the six
  segments. The highest-observability segment, with $\lambda_{\min}[J^{(m)}] \approx 8.21 \times 10^4$
  km$^{-2}$, has the lowest median entropy, 1.64 nats ($\bar{H}_t = 0.18$). The remaining five segments
  have $9.39 \times 10^3 \leq \lambda_{\min}[J^{(m)}]/(1\,{\rm km^{-2}}) \leq 3.23 \times 10^4$. Their median entropies
  lie between 2.24 and 2.47 nats ($0.24 \leq \bar{H}_t \leq 0.27$). Although based on only six segments and therefore illustrative
  rather than statistically conclusive, this trend is consistent with the expectation that the least
  informative spatial direction over the accumulation window influences the posterior spread. Third,
  the right panel exhibits a positive correlation between $H_t$ and $e_{\rm pos}(t)$, with $\rho\big[H_t, e_{\rm pos}(t)\big] =
  0.30$. The position error spans $1.3$ to $82.7$ m, with a median of $23.5$ m. Timesteps at which
  the algorithm is uncertain (high $H_t$) tend to be the same timesteps at which the position estimate
  is furthest from truth. The entropy $H_t$ depends on the spread of $\gamma_t$ and not on its location. At the ten timesteps with $e_{\rm pos}(t) > 2\,\Delta_{\rm cell}$, where $\gamma_t$ is concentrated on a single displaced mode (Section~\ref{SubSec:MagNavError}), $H_t$ ranges from the 38th percentile of its values over the flight to its maximum. The foregoing correlation supports the interpretation of $H_t$ as a truth-free diagnostic of posterior concentration and map-conditioned ambiguity. The single-realisation correlations reported here are corroborated by the ensemble results of Section~\ref{Sec:Performance}.

\section{Performance Tests}\label{Sec:Performance}

The results of Section~\ref{Sec:Validation} correspond to a single realisation of the stochastic processes governing the INS error model (Table~\ref{tab:ins_params}) and the magnetic measurement noise (Table~\ref{tab:mag_params}). In this section, we use two Monte Carlo experiments, each comprising $10^3$ independent trials, to measure (i) the sensitivity of the navigation errors to the combined sensor-error realisation for a fixed flight path; and (ii) the variation of the navigation errors across flight paths over different regions of the map, i.e.\ trajectories that sample different magnetic anomaly structure. In Section~\ref{SubSec:Noise}, we hold the flight path fixed and draw the INS and magnetometer error processes independently in each trial. In Section~\ref{SubSec:Observability}, we hold the sensor-error realisation fixed and vary the initial aircraft heading $\psi$ and start location, generating an ensemble of distinct flight paths over the \texttt{Renfrew\_395} map. Section~\ref{SubSec:Comparison} compares the factor graph with the forward-only PMF, the EKF, the particle filter, and the Viterbi path on the two ensembles. Section~\ref{SubSec:EntropyFlag} tests a threshold on $H_t$ as a flag for segments with large position errors.

\subsection{Sensitivity to sensor-error realisations}\label{SubSec:Noise}

In this experiment, we hold the flight path and factor graph design fixed to the representative realisation of Section~\ref{Sec:Validation} and draw $10^3$ independent realisations of the INS and magnetometer error processes. All other parameters, including the map, the trajectory, and the factor graph configuration ($n = 101$, $T = 40$, $M = 6$), are identical across trials. We estimate the noise parameters $\sigma_{\rm mag}$ and $\sigma_v$ using Equations~(\ref{eq:debias})--(\ref{eq:evidence_product}) for each sensor-error realisation, whose median values are $\hat{\sigma}_{\rm mag} = 0.91$~nT ($0.80$--$1.15$~nT) and $\hat{\sigma}_v = 5.29$~m\,s$^{-1}$ ($3.63$--$6.56$~m\,s$^{-1}$), and the ranges in brackets correspond to the 16th--84th percentile ranges.

\begin{figure}
    \centering
    \includegraphics[width=1.0\textwidth]{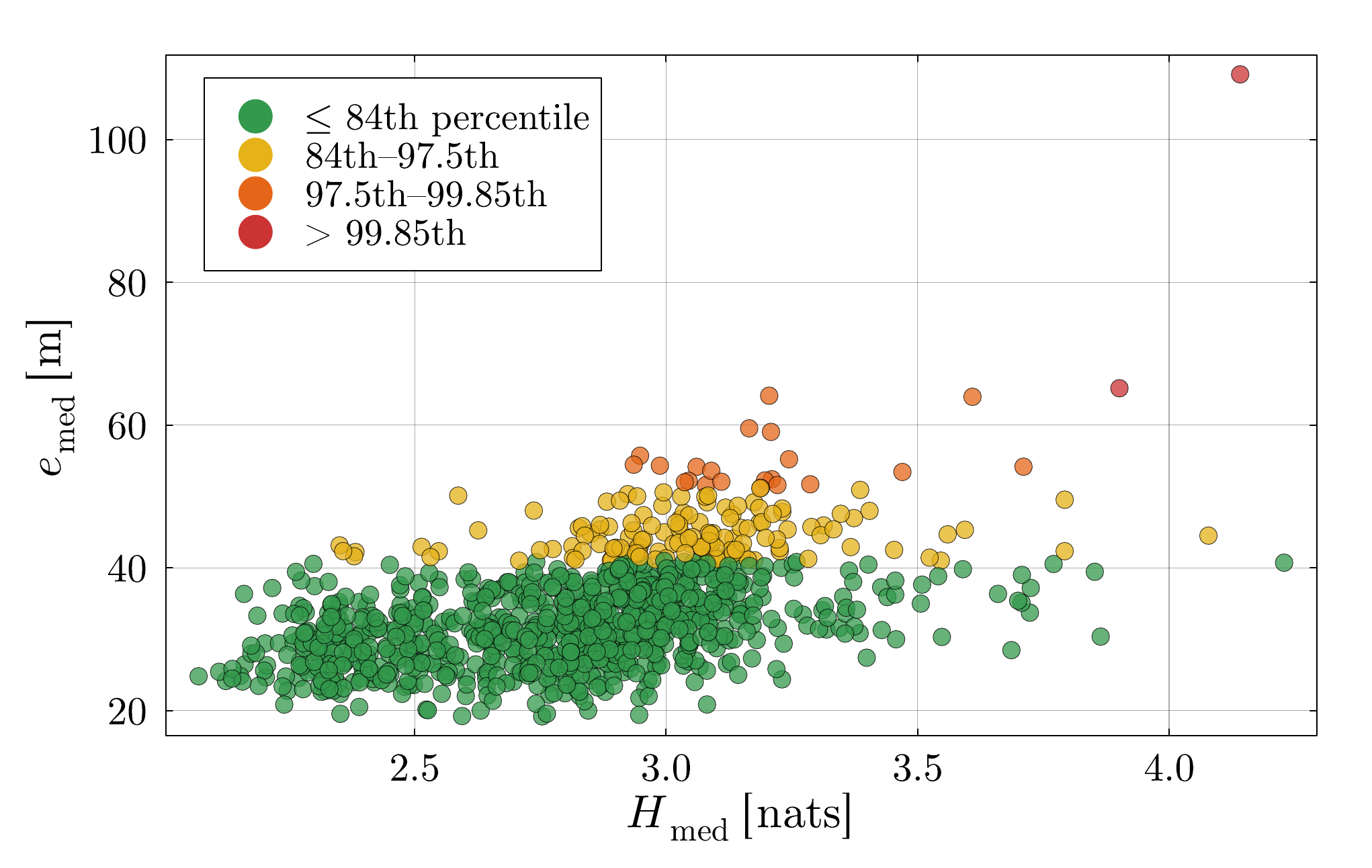}
    \caption{Per-trial median posterior entropy $H_{\rm med} = \operatorname{median}_t H_t$
  [Equation~(\ref{eq:entropy}); units: nats] versus per-trial median position error $e_{\rm med} = \operatorname{median}_t e_{{\rm pos},t}$ (units: m) for $10^3$ independent realisations of the INS and magnetometer error
  processes on a fixed flight path (Section~\ref{SubSec:Scenario}). Each point corresponds to a single sensor-error realisation, with the medians taken over all $N_{\rm FG} = 240$ timesteps within each trial. Points are coloured by percentile bands of $e_{\rm med}$ across trials: green points denote trials below the 84th percentile ($e_{\rm med} \leq 40.9$~m), yellow points denote trials between the 84th and 97.5th percentiles ($40.9$--$50.6$~m), orange points denote trials between the 97.5th and 99.85th percentiles ($50.6$--$64.6$~m), and red points denote trials above the 99.85th percentile ($>64.6$~m). Both axes are plotted on linear scales. The factor graph configuration
  ($n = 101$, $T = 40$, $M = 6$) is identical across all trials}
    \label{Fig:Noise}
\end{figure}

In Figure~\ref{Fig:Noise}, we plot the per-trial median posterior entropy, $H_{\rm med} = \operatorname{median}_t H_t$ (units: nats), on the horizontal axis against the per-trial median position error, $e_{\rm med} = \operatorname{median}_t e_{{\rm pos},t}$ (units: m), on the vertical axis. Specifically, each point corresponds to a single sensor-error realisation, with the medians taken over all $N_{\rm FG} = 240$ timesteps within that trial. The points are coloured by percentile bands of $e_{\rm med}$ across trials: green denotes trials below the 84th percentile ($e_{\rm med} \leq 40.9$ m), yellow denotes trials between the 84th and 97.5th percentiles ($40.9$--$50.6$ m), orange denotes trials between the 97.5th and 99.85th percentiles ($50.6$--$64.6$ m), and red denotes trials above the 99.85th percentile ($>64.6$ m). Both axes are plotted on linear scales.

The results in Figure~\ref{Fig:Noise} have four key features. First, across the $10^3$ trials, $e_{\rm med}$ has a median of $32.7$ m and a 16th--84th percentile range of $26.6$--$40.9$ m. For comparison, the same quantity for the INS (not plotted) has a median of $362$ m and a 16th--84th percentile range of $186$--$599$ m. That is, forward-backward inference yields a $91\%$ relative improvement when compared with INS drift. Second, the analysis reveals that $19.2 \lesssim e_{\rm med}/(1 \, \rm{m}) \lesssim 109.1$, indicating that the dispersion associated with $e_{\rm med}$ for a fixed flight path is moderately sensitive to the particular sensor-error realisation. The majority of trials (840 out of 1000) fall below the 84th-percentile threshold. Only 2 trials exceed the 99.85th-percentile threshold, as expected. Third, the correlation between $H_{\rm med}$ and $e_{\rm med}$ is moderate, with $\rho(H_{\rm med}, e_{\rm med}) = 0.49$, indicating that, when the flight path is held fixed, the posterior entropy captures part of the inter-trial variation in navigation error. Pooled over all timesteps and trials, the median of $H_t$ is $2.79$ nats ($\bar{H}_t = 0.30$) [Equation~(\ref{eq:entropy_norm})]. Fourth, the worst-performing trials tend to cluster at elevated entropy rather than being scattered uniformly across the entropy range. All 25 trials above the 97.5th percentile of $e_{\rm med}$ (orange and red points) lie in the upper half of the $H_{\rm med}$ distribution. Of the 160 trials above the 84th percentile, 87\% lie in the upper half, compared with 43\% of the remaining trials. This suggests that $H_t$ provides some coarse separation of the error tail even when the geometry is fixed. We quantify the foregoing separation at the segment level in Section~\ref{SubSec:EntropyFlag}.

\subsection{Observability and entropy across flight paths}\label{SubSec:Observability}

In this experiment, we fix the random draws of the sensor-error processes and the factor graph design ($n = 101$, $T = 40$, $M = 6$)
  and vary the initial heading $\psi$ and start location, generating $10^3$ distinct flight paths over
  the \texttt{Renfrew\_395} map. The heading is measured anticlockwise from east, i.e.\ $\psi = 90^\circ$ corresponds to north. It is sampled uniformly within $45^\circ$ of a reference heading of approximately $95^\circ$, and the start
  location is drawn randomly within the map domain. Draws for which the flight path leaves the map are rejected and redrawn, which restricts the accepted headings to $69.3^\circ \leq \psi \leq 121.4^\circ$, i.e.\ bearings within approximately $30^\circ$ of north. Each trajectory follows the same sinusoidal
  lateral profile ($N_{\rm waves} = 3$ cycles) at the same airspeed and altitude as in Section~\ref{Sec:Validation}, but
  traverses a different region of the anomaly field. Because the sensor-error realisation is common to all
  trials, the variation in navigation performance across trials reflects the map structure and trajectory geometry encountered along each flight path. We estimate the noise parameters $\sigma_{\rm mag}$ and $\sigma_v$ using Equations~(\ref{eq:debias})--(\ref{eq:evidence_product}) for each flight path, whose median values are $\hat{\sigma}_{\rm mag} = 0.94$~nT ($0.86$--$1.07$~nT) and $\hat{\sigma}_v = 3.17$~m\,s$^{-1}$ ($2.67$--$4.09$~m\,s$^{-1}$), and the ranges in brackets correspond to the 16th--84th percentile ranges.

  \begin{figure}
  \centering
  \includegraphics[width=1.0\textwidth]{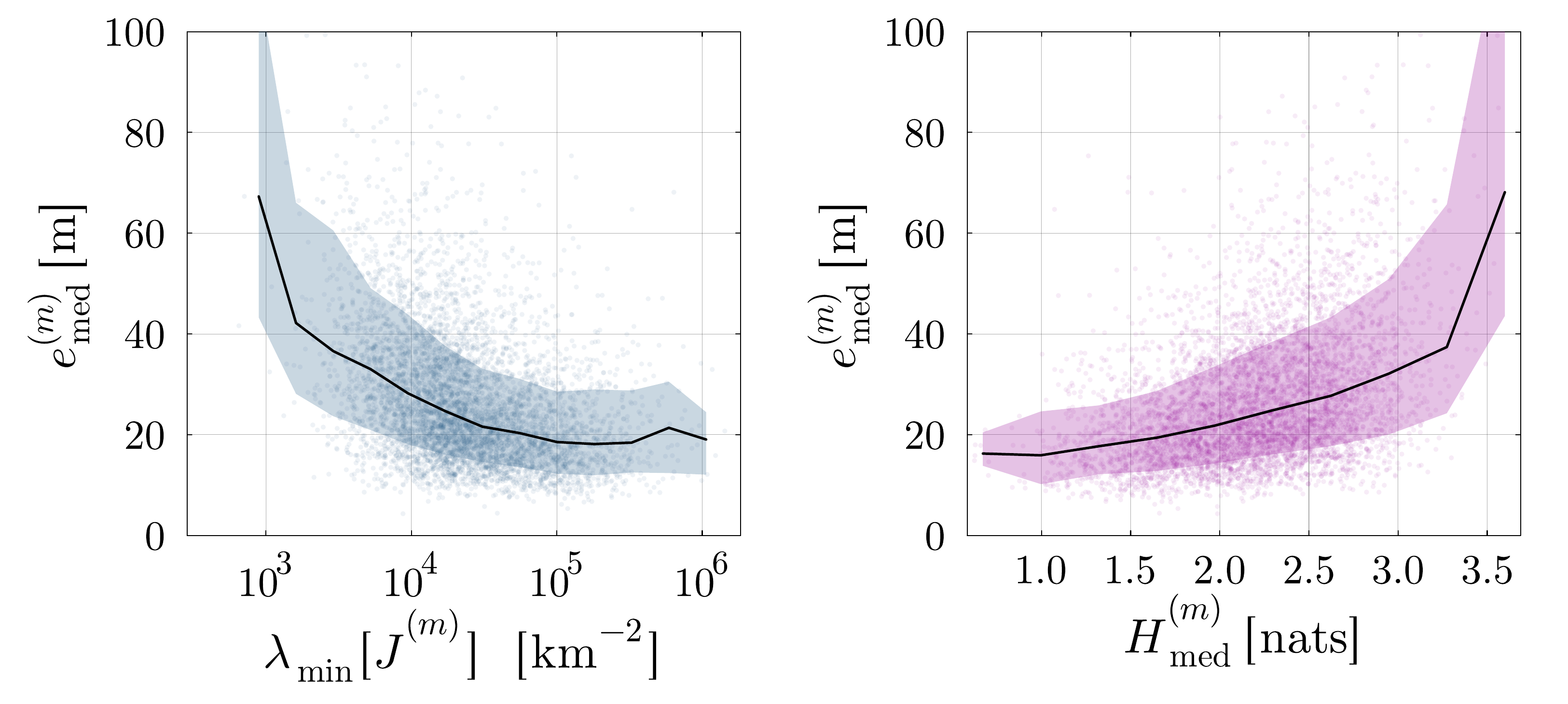}
  \caption{Within-segment median position error $e^{(m)}_{\rm med} = \operatorname{median}_{t \in \mathcal{T}^{(m)}} e_{{\rm pos},t}$ (units: m) for
  $10^3$ distinct flight paths over the \texttt{Renfrew\_395} map with a common sensor-error
  realisation. Each trajectory contributes $M = 6$ segments, yielding $6 \times 10^3$
  segment-level samples in total. In both panels, individual segment observations are
  overlaid as faint markers, the solid black curve traces the binwise conditional median,
  and the shaded band spans the 16th to 84th percentile range within each bin. Left panel:
  the horizontal axis is the segment-level minimum eigenvalue of the accumulated Fisher
  information $\lambda_{\min}[J^{(m)}]$ (units: km$^{-2}$), plotted on a logarithmic scale.
  Right panel: the horizontal axis is the within-segment median posterior entropy
  $H^{(m)}_{\rm med} = \operatorname{median}_{t \in \mathcal{T}^{(m)}} H_t$ [Equation~(\ref{eq:entropy}); units: nats], plotted
  on a linear scale. The factor graph configuration ($n = 101$, $T = 40$, $M = 6$) is identical
  across all trials; see Table~\ref{tab:fg_params}}
  \label{Fig:MCRealizations}
  \end{figure}

  In Figure~\ref{Fig:MCRealizations}, we plot the within-segment median position error, $e^{(m)}_{\rm med} = \operatorname{median}_{t \in \mathcal{T}^{(m)}} e_{{\rm pos},t}$ (units: m), against two diagnostic quantities, viz.\ the minimum eigenvalue $\lambda_{\min}[J^{(m)}]$ of the per-segment Fisher information (units: km$^{-2}$; left panel) and the within-segment median posterior entropy $H^{(m)}_{\rm med}$ (units: nats; right panel). Each of the $10^3$ trajectories contributes $M =
  6$ segments, yielding $6 \times 10^3$ segment-level samples in total. The minimum eigenvalue $\lambda_{\min}[J^{(m)}]$ measures the least-informative horizontal direction within a segment. Specifically, a segment with large gradient magnitude but little directional diversity can have large trace but poor two-dimensional position observability. The horizontal axis is plotted on a logarithmic scale in the left panel and on a linear scale in the right panel. In both panels, the individual segment observations are overlaid as faint
  markers, the solid black curve traces the binwise conditional median, and the shaded band spans the
  16th to 84th percentile range within each bin.

  The left panel has three notable features. First, the binwise median segment error shows an
  overall decrease with increasing observability, falling from approximately $67$ m at $\lambda_{\min}
  [J^{(m)}] \approx 9 \times 10^2$ km$^{-2}$ to approximately $18$ m at $\lambda_{\min}[J^{(m)}] \approx 1.8
  \times 10^5$ km$^{-2}$. At the segment level, $\rho\big\{\log_{10}\lambda_{\min}[J^{(m)}], e^{(m)}_{\rm med}\big\} = -0.33$. The decrease in segment error with increasing $\lambda_{\min}[J^{(m)}]$ is consistent with the qualitative Cram\'{e}r--Rao expectation that larger accumulated information permits smaller position uncertainty. Second, the conditional dispersion (width of the shaded band) narrows as observability
  increases: segments with strong magnetic gradients in all spatial directions produce not only lower
  median errors but also tighter error distributions. Third, the binwise median flattens to $18 \lesssim e^{(m)}_{\rm med}/(1 \, \rm{m}) \lesssim 21$ for $\lambda_{\min}[J^{(m)}] \gtrsim 10^5$ km$^{-2}$.

The right panel shows the complementary view. First, the binwise median error rises overall with
increasing entropy, from approximately $16 \, \rm{m}$ at $H^{(m)}_{\rm med} \approx 0.7$
  nats ($\bar{H}_t \approx 0.07$) to approximately $68 \, \rm{m}$ at $H^{(m)}_{\rm med} \approx 3.6$ nats ($\bar{H}_t \approx 0.39$). At the segment level, $\rho\big[H^{(m)}_{\rm med}, e^{(m)}_{\rm med}\big] = 0.38$. The positive entropy-error correlation indicates that $H_t$ is informative about map-conditioned ambiguity, but it should not be interpreted as a calibrated predictor of realised error. Pooled over all timesteps and trajectories, the median of $H_t$ is $2.23$ nats ($\bar{H}_t = 0.24$). Second, the
  conditional dispersion broadens with increasing entropy: high-entropy segments are associated with
  larger typical errors and greater variability in realised performance. This is consistent with
  the physical interpretation that a diffuse posterior reflects weak map-matching constraints,
  suggesting that the position estimate becomes more sensitive to the sensor errors.
  Third, comparing the two panels, the correlations of the two diagnostics with the segment error are of similar magnitude, i.e.\ $|\rho| = 0.38$ for $H^{(m)}_{\rm med}$ and $|\rho| = 0.33$ for $\log_{10}\lambda_{\min}[J^{(m)}]$. The correlation between the two diagnostics is $\rho\big\{H^{(m)}_{\rm med}, \log_{10}\lambda_{\min}[J^{(m)}]\big\} = -0.75$. The foregoing correlations support the interpretation of $H_t$ as a truth-free diagnostic of map-conditioned ambiguity (Section~\ref{SubSec:ShannonEntropy}).

\subsection{Comparison with other estimators}\label{SubSec:Comparison}

In this experiment, we compare the factor graph with four other estimators on the two ensembles of Sections~\ref{SubSec:Noise} and~\ref{SubSec:Observability}: (i) the forward-only PMF; (ii) the EKF; (iii) the bootstrap particle filter with $10^4$ and $10^5$ particles; and (iv) the Viterbi path. The forward-only PMF is the forward sweep of the factor graph (Section~\ref{SubSec:forward_backward}), run with the same grids, pruning, parameters, and measurements. Its estimate at $t_k$ is the MMSE estimate of Equation~(\ref{eq:mmse}), with $\gamma_{t_k}$ replaced by the normalised forward variable $\alpha_{t_k}$. The EKF and the particle filter operate on the same two-dimensional model, with the same parameters and measurements, and the Viterbi path is computed on the same grids. The four estimators are specified in Appendix~\ref{App:Viterbi}. In Table~\ref{tab:estimators}, we list the median position error of each estimator, pooled over all timesteps and trials, and the run time per timestep on one core. The marginal MAP estimate of Equation~(\ref{eq:map}) is read off the same posterior $\gamma_{t_k}$ as the MMSE estimate.

\begin{table}[t]
\centering
\caption{Median position error of each estimator, pooled over all timesteps and trials, for the $10^3$ sensor-error realisations of Section~\ref{SubSec:Noise} and the $10^3$ flight paths of Section~\ref{SubSec:Observability}, and run time per timestep on one core. The EKF and the particle filter use the same two-dimensional model, parameters, and measurements as the factor graph. The PMF, the Viterbi path, and the factor graph use the same grids and pruning.}
\label{tab:estimators}
\begin{tabular}{lccc}
\toprule
 & \multicolumn{2}{c}{{\bfseries Median error (m)}} & \\
\cmidrule(lr){2-3}
{\bfseries Estimator} & {\bfseries Sensor-error realisations} & {\bfseries Flight paths} & {\bfseries Run time (s)} \\
\midrule
Factor graph, MMSE         & 32.8 & 23.8 & 0.022--0.032 \\
Factor graph, marginal MAP & 35.5 & 27.0 & 0.022--0.032 \\
Viterbi path               & 38.3 & 30.5 & 0.019--0.022 \\
PMF (forward only)         & 57.7 & 41.9 & 0.013--0.020 \\
EKF                        & 57.9 & 41.6 & $<10^{-4}$ \\
Particle filter, $10^4$    & 57.6 & 41.6 & 0.0013 \\
Particle filter, $10^5$    & 57.5 & 41.5 & 0.013 \\
\bottomrule
\end{tabular}
\end{table}

The results in Table~\ref{tab:estimators} have four key features. First, across sensor-error realisations, the factor graph reduces the median position error from $57.7$ m for the PMF to $32.8$ m, i.e.\ by $43\%$. Across flight paths, it reduces the median position error from $41.9$ m to $23.8$ m, also by $43\%$. Because the two estimators share the grids, pruning, parameters, and measurements, the reduction is due to the backward sweep, which incorporates the later measurements $z_{t_{k+1}:t_T}$ of each segment through $\beta_{t_k}$. Second, the three causal filters, i.e.\ the PMF, the EKF, and the particle filter, agree. Their median errors lie within $0.4$ m of one another on each ensemble. They range from $57.5$ m to $57.9$ m across sensor-error realisations and from $41.5$ m to $41.9$ m across flight paths. Reducing the number of particles from $10^5$ to $10^4$ changes the median error by $0.1$ m. The PMF approximates the filtering density $p(\bm{s}_{t_k} \mid z_{t_1:t_k})$ on a grid, and the particle filter approximates the same density by samples. Third, the Viterbi path, computed from the same measurements $z_{\mathcal{T}}$ as the factor graph, reaches median errors of $38.3$ m for the sensor-error realisations and $30.5$ m for the flight paths. In per-trial median error, the Viterbi path exceeds the MMSE estimate by $5.3$ m for the sensor-error realisations and by $6.6$ m for the flight paths. The marginal MAP estimate exceeds the MMSE estimate by $2.5$ m for the sensor-error realisations and by $3.1$ m for the flight paths. The difference for the marginal MAP estimate bounds the cost of quantisation (Section~\ref{SubSec:Idealisations}), so quantisation accounts for at most about half of the gap to the Viterbi path. The two cell-centre estimates differ in how they read the posterior. The Viterbi path is the most probable sequence of cells over a segment, and the marginal MAP estimate is the most probable cell at each timestep. Fourth, on one core, the factor graph takes $0.022$--$0.032$ s per timestep. The particle filter with $10^5$ particles takes $0.013$ s per timestep, and the EKF less than $10^{-4}$ s.

\subsection{Entropy as a threshold flag}\label{SubSec:EntropyFlag}

In this experiment, we test a threshold $\kappa$ on the posterior entropy as a flag for segments with large position errors, using the $6 \times 10^3$ segments of each of the ensembles of Sections~\ref{SubSec:Noise} and~\ref{SubSec:Observability}. We flag segment $m$ when $H^{(m)}_{\rm med} > \kappa$. To score the flag, we label segment $m$ as a large-error segment when $e^{(m)}_{\rm med} > e_{\rm lim}$, where $e_{{\rm pos},t}$ is the position error of the MMSE estimate of the factor graph and $e_{\rm lim}$ is an error limit. The flag depends on $H_t$ only. Let $P_{\rm D}(\kappa)$ denote the detection rate, i.e.\ the fraction of large-error segments that are flagged, and $P_{\rm FA}(\kappa)$ the false-alarm rate, i.e.\ the fraction of the remaining segments that are flagged. We consider three error limits on each ensemble: (i) $e_{\rm lim} = 40.9$ m, the 84th percentile of $e_{\rm med}$ across sensor-error realisations (Section~\ref{SubSec:Noise}); (ii) $e_{\rm lim} = 2\Delta_{\rm cell} = 58.5$ m; and (iii) $e_{\rm lim} = 3\Delta_{\rm cell} = 87.8$ m. Varying $\kappa$ from the smallest to the largest value of $H^{(m)}_{\rm med}$ across segments traces the receiver operating characteristic (ROC) curve of $P_{\rm D}(\kappa)$ against $P_{\rm FA}(\kappa)$, whose area under the curve (AUC) equals 0.5 for a flag independent of the error and 1 for a flag that separates the large-error segments from the rest completely. We repeat the test, with the same large-error labels, using the entropy of the forward pass, i.e.\ Equation~(\ref{eq:entropy}) with $\gamma_{t_k}$ replaced by the normalised forward variable $\alpha_{t_k}$, which depends on the measurements $z_{t_1:t_k}$ up to $t_k$.

\begin{figure}
    \centering
    \includegraphics[width=1.0\textwidth]{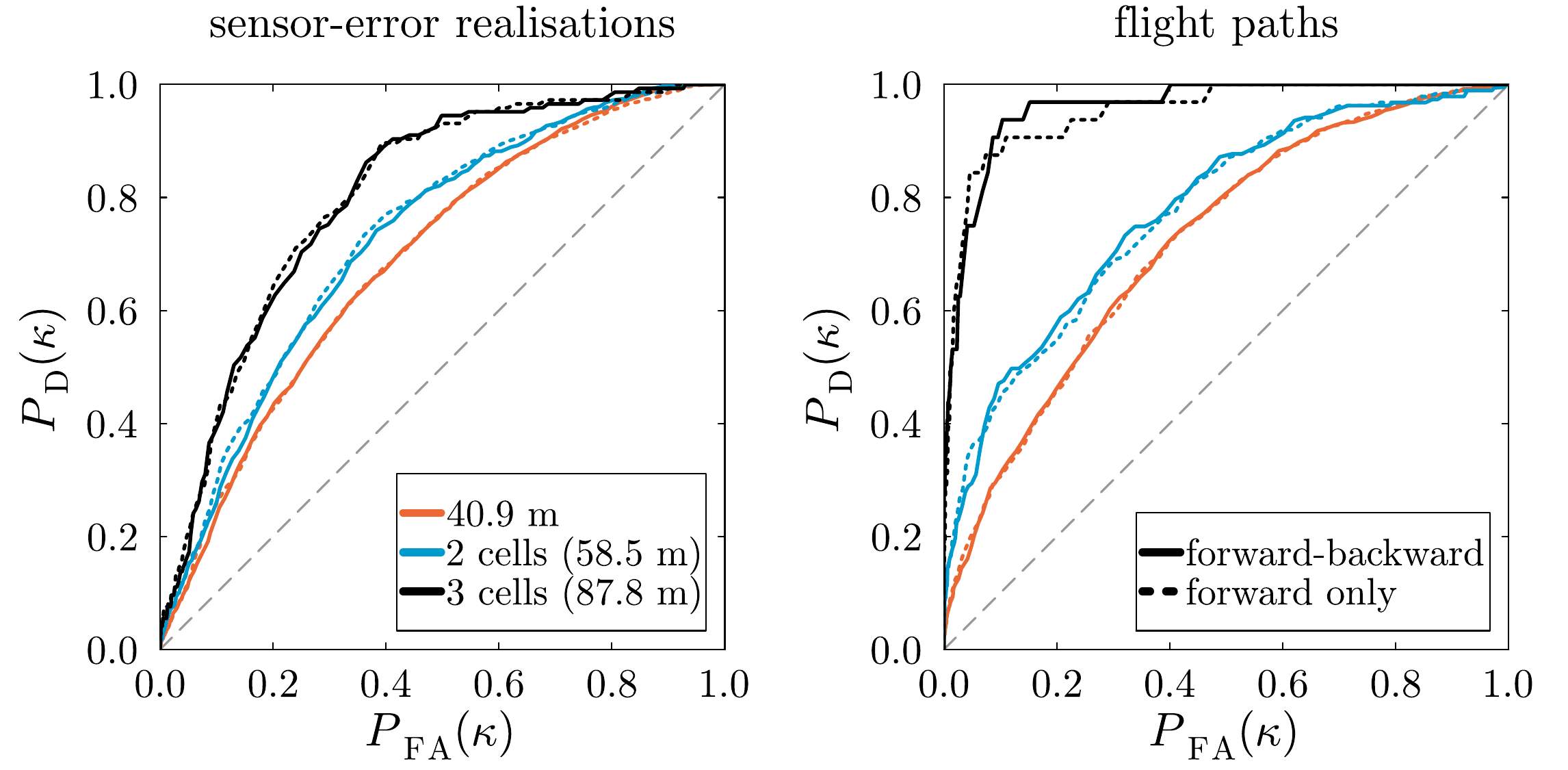}
    \caption{Detection rate $P_{\rm D}(\kappa)$ versus false-alarm rate $P_{\rm FA}(\kappa)$ of a threshold $\kappa$ on the within-segment median posterior entropy $H^{(m)}_{\rm med} = \operatorname{median}_{t \in \mathcal{T}^{(m)}} H_t$ [Equation~(\ref{eq:entropy}); units: nats], for the $6 \times 10^3$ segments of the $10^3$ sensor-error realisations of Section~\ref{SubSec:Noise} (left panel) and the $10^3$ flight paths of Section~\ref{SubSec:Observability} (right panel). A segment has a large error when its median position error $e^{(m)}_{\rm med}$ exceeds $40.9$ m (orange curves), two cells, i.e.\ $58.5$ m (blue curves), or three cells, i.e.\ $87.8$ m (black curves). Solid curves: entropy of the posterior $\gamma_{t_k}$ of the factor graph. Dotted curves: entropy of the normalised forward variable $\alpha_{t_k}$. The grey, dashed diagonal corresponds to a flag independent of the error}
    \label{Fig:EntropyROC}
\end{figure}

In Figure~\ref{Fig:EntropyROC}, we plot $P_{\rm D}(\kappa)$ against $P_{\rm FA}(\kappa)$ for the sensor-error realisations (left panel) and the flight paths (right panel). The solid curves use the entropy of the posterior $\gamma_{t_k}$ of the factor graph, and the dotted curves the entropy of the forward pass. The curves are coloured by error limit: $40.9$ m (orange), two cells (blue), and three cells (black). The grey, dashed diagonal corresponds to a flag independent of the error.

The results in Figure~\ref{Fig:EntropyROC} have three key features. First, for $e_{\rm lim} = 2\,\Delta_{\rm cell}$, the AUC is $0.73$ across sensor-error realisations. At $P_{\rm FA} = 0.10$, the flag detects $26\%$ of the large-error segments, at $\kappa = 3.27$ nats ($\kappa/\ln N_s = 0.35$). Across flight paths, the AUC is $0.78$. At $P_{\rm FA} = 0.10$, the flag detects $47\%$ of the large-error segments, at $\kappa = 2.77$ nats ($\kappa/\ln N_s = 0.30$). The thresholds differ between the ensembles because the distribution of $H_t$ differs. Its pooled median is $2.79$ nats ($\bar{H}_t = 0.30$) across sensor-error realisations and $2.23$ nats ($\bar{H}_t = 0.24$) across flight paths (Sections~\ref{SubSec:Noise} and~\ref{SubSec:Observability}). Second, the AUC increases with the error limit. Across sensor-error realisations, it rises from $0.69$ for $e_{\rm lim} = 40.9$ m to $0.80$ for $e_{\rm lim} = 3\,\Delta_{\rm cell}$, which 145 segments exceed. Across flight paths, it rises from $0.72$ to $0.96$, and 32 segments exceed $3\,\Delta_{\rm cell}$. At $P_{\rm FA} = 0.10$, the fraction of large-error segments detected rises from $24\%$ to $39\%$ across sensor-error realisations, and from $31\%$ to $94\%$ across flight paths. Third, the entropy of the forward pass gives the same AUC to within $0.01$ at every error limit on each ensemble (dotted curves). Accordingly, the flag applies equally to the forward-only PMF, while the backward sweep reduces the position error (Section~\ref{SubSec:Comparison}).

\section{Conclusion}\label{Sec:Conclusion}
In this paper, we formulate airborne magnetic anomaly navigation as inference on a chain-structured factor graph and solve it recursively using the forward-backward algorithm. The method discretises the aircraft position domain onto a map-aligned grid of $N_s = n \times n$ candidate cells at each timestep, divides the flight into $M$ non-overlapping segments of $T$ timesteps, recentres the grid on the estimate at every segment boundary, and computes hidden state marginal posteriors within each segment. The approach yields the MMSE position estimate [Equation~(\ref{eq:mmse})] and the Shannon posterior entropy $H_t$ [Equation~(\ref{eq:entropy})] at each timestep as closed-form outputs of the inference. The forward sweep of the factor graph is the PMF (Section~\ref{SubSec:forward_backward}), which represents non-Gaussian or multimodal posteriors explicitly on a discrete grid, while retaining deterministic dynamic-programming recursions for a fixed candidate support. The backward sweep conditions each estimate on the later measurements of its segment. The MMSE estimate is a posterior-weighted average over the candidate grid, i.e.\ the estimate is not constrained to grid cell centres. The posterior entropy $H_t$ provides a truth-free diagnostic of posterior concentration and map-conditioned ambiguity. The entropy $H_t$ reflects multimodal ambiguity on the discrete grid. It is computed directly from the marginal posterior $\gamma_t$ over the candidate positions.

The results are organised in two parts. In Section~\ref{Sec:Validation}, a representative test case ($v = 68$ m s$^{-1}$, $h = 1000$ m, $t_{\rm flight} = 1200$ s, $n = 101$, $T = 40$, $M = 6$) yields a median MMSE position error of $23.5$ m. The largest MMSE error is $82.7$ m. For the uncorrected INS, the median position error is $284.9$ m and the final error is $\approx 1074$ m. The posterior entropy $H_t$ correlates negatively with the per-timestep Fisher trace, with $\rho\big[H_t, \log_{10}\operatorname{tr}(J_t)\big] = -0.80$. It correlates positively with the realised position error, with $\rho\big[H_t, e_{\rm pos}(t)\big] = 0.30$. In Section~\ref{Sec:Performance}, two Monte Carlo experiments ($10^3$ trials each) probe the robustness of these findings, with the noise parameters estimated in each trial. In each experiment, the recentred grid contains the true position at every timestep (Section~\ref{SubSec:SelectingT}). Across sensor-error realisations (Section~\ref{SubSec:Noise}), the per-trial median position error $e_{\rm med}$ has a median of $32.7$ m and a 16th--84th percentile range of $26.6$--$40.9$ m. This is a $91\%$ reduction relative to the median INS error of $362$ m. The correlation between the per-trial median entropy $H_{\rm med}$ and $e_{\rm med}$ is moderate, with $\rho(H_{\rm med}, e_{\rm med}) = 0.49$. Across flight paths (Section~\ref{SubSec:Observability}), where the heading and start location are varied over the map, the binwise median segment error decreases from approximately $67$ m at $\lambda_{\min}[J^{(m)}] \approx 9 \times 10^2$ km$^{-2}$ to approximately $18$ m at $\lambda_{\min}[J^{(m)}] \approx 1.8 \times 10^5$ km$^{-2}$, consistent with the qualitative Cram\'{e}r--Rao expectation that larger accumulated information permits smaller position uncertainty. At the segment level, $\rho\big[H^{(m)}_{\rm med}, e^{(m)}_{\rm med}\big] = 0.38$ and $\rho\big\{\log_{10}\lambda_{\min}[J^{(m)}], e^{(m)}_{\rm med}\big\} = -0.33$. The correlation between the two diagnostics is $\rho\big\{H^{(m)}_{\rm med}, \log_{10}\lambda_{\min}[J^{(m)}]\big\} = -0.75$.

Sections~\ref{SubSec:Comparison} and~\ref{SubSec:EntropyFlag} quantify the accuracy of the factor graph against four baseline estimators and the performance of $H_t$ as a flag for segments with large position errors. Across sensor-error realisations, the factor graph reduces the median position error from $57.7$ m for the PMF to $32.8$ m, i.e.\ by $43\%$. Across flight paths, it reduces the median position error from $41.9$ m to $23.8$ m, also by $43\%$. The EKF and the particle filter lie within $0.4$ m of the PMF, so the gain over the causal filters is due to the backward sweep. The forward sweep yields an estimate at each timestep as its measurement arrives. The backward sweep refines it after a delay of at most $T\,\Delta t = 200$ s. On one core, the factor graph takes $0.022$--$0.032$ s per timestep, i.e.\ under $1\%$ of $\Delta t = 5$ s. The particle filter with $10^5$ particles takes $0.013$ s per timestep, and the EKF less than $10^{-4}$ s (Table~\ref{tab:estimators}). A threshold on $H^{(m)}_{\rm med}$ flags the segments whose median error exceeds two cells. Across sensor-error realisations, the AUC of this flag is $0.73$, and the flag detects $26\%$ of these segments at $P_{\rm FA} = 0.10$. Across flight paths, the AUC is $0.78$, and the flag detects $47\%$ of these segments at $P_{\rm FA} = 0.10$. For an error limit of three cells, the AUC rises to $0.80$ across sensor-error realisations and $0.96$ across flight paths.

There are at least four ways in which the present work can be extended. First, the factor graph framework can be applied to real flight data, replacing the synthetic INS and magnetometer signals with measurements from an operational platform, and estimating the bias and noise parameters of Section~\ref{SubSec:DataPrep} from the segments already flown. Second, the hidden state vector can be augmented to include additional estimated quantities, such as the FOGM magnetic bias $S(t)$, by attaching further variable nodes to the existing chain, and further sensor modalities can enter as additional factors without modification to the inference recursion. One such quantity is a map-error state correlated along the track, which the white-noise measurement factor of Equation~(\ref{eq:meas_magnav}) does not model (see Appendix~\ref{App:MapMismatch}). Third, the candidate grid can be extended to $n_z$ altitude levels [Section~\ref{SubSec:SimIdealisations}, idealisation (i)], which multiplies the transition cost by up to $n_z^2$, so that a grid with ten levels takes at most $3.2$ s per timestep on one core, within $\Delta t = 5$ s. Fourth, the posterior entropy $H_t$ and the minimum eigenvalue of the accumulated Fisher information $\lambda_{\min}[J^{(m)}]$ can inform flight planning strategies by identifying, prior to flight, map regions and heading configurations that maximise magnetic observability and minimise expected navigation error.

\section*{Competing interests} None.

\begin{appendices}

\section{Baseline estimators}\label{App:Viterbi}

The forward-backward algorithm of Section~\ref{SubSec:forward_backward} yields the marginal posterior $\gamma_{t_k}$ at each timestep, from which the factor graph obtains the MMSE estimate of Equation~(\ref{eq:mmse}). By way of comparison, we consider four baseline estimators, viz.\ the forward-only PMF, the EKF, the bootstrap particle filter, and the Viterbi path. All four are compared with the factor graph in Section~\ref{SubSec:Comparison}, and three of them also appear in Figure~\ref{Fig:ErrorTracking}. Each baseline operates on the two-dimensional position model of Section~\ref{Sec:Magnav}, with the same motion model (Section~\ref{SubSec:Trans}), observation model (Section~\ref{SubSec:Meas}), debiased measurements [Equation~(\ref{eq:debias})], and noise parameters [Equation~(\ref{eq:evidence_max})] as the factor graph, so that the estimators differ only in the inference. In this appendix, we specify the four baselines. The forward-only PMF is described in Appendix~\ref{App:PMF}. The EKF and the bootstrap particle filter are specified in Appendices~\ref{App:EKF} and~\ref{App:PF}, respectively. The Viterbi path, and the cost of restricting an estimate to cell centres, are discussed in Appendix~\ref{App:ViterbiPath}.

\subsection{Point-mass filter}\label{App:PMF}

The forward-only PMF \citep[Section~2.3]{Bergman_1999} is the forward sweep of Section~\ref{SubSec:forward_backward}, run with the same pruning as the factor graph. Its position estimate at time $t_k$ is given by Equation~(\ref{eq:mmse}), with $\gamma_{t_k}(j)$ replaced by the normalised forward variable $\alpha_{t_k}[j]$. At the last timestep of each segment, $\beta_{t_T}[j] = 1$ [Equation~(\ref{eq:beta_init})], so the PMF and the factor graph return the same estimate and the same terminal posterior. Accordingly, the carried correction of Equation~(\ref{eq:carried_correction}) and the prior of Equation~(\ref{eq:prior_carryover}) are common to the two estimators, and the two run on identical candidate grids.

\subsection{Extended Kalman filter}\label{App:EKF}

The state of the EKF is the horizontal position $\bm{s}_{t_k}$ in local north-east coordinates, with covariance $\bm{\Sigma}_{t_k} \in \mathbb{R}^{2 \times 2}$. INS error states are not estimated, so that the EKF and the factor graph share the model of Section~\ref{Sec:Magnav}. The prediction step propagates the estimate through the motion model of Equation~(\ref{eq:motion_model}), viz.\
\begin{equation}
\hat{\bm{s}}^{-}_{t_k} = \hat{\bm{s}}_{t_{k-1}} + \bm{v}_{t_{k-1}}\,\Delta t, \qquad
\bm{\Sigma}^{-}_{t_k} = \bm{\Sigma}_{t_{k-1}} + \sigma_v^2\,\Delta t^2\,\bm{I}_2.
\label{eq:ekf_predict}
\end{equation}
The update step linearises the observation model of Equation~(\ref{eq:obs_model}) about the predicted position, viz.\
\begin{equation}
\begin{aligned}
\bm{k}_{t_k} &= \frac{\bm{\Sigma}^{-}_{t_k}\,\bm{g}_{t_k}}{\bm{g}_{t_k}^{\rm T}\,\bm{\Sigma}^{-}_{t_k}\,\bm{g}_{t_k} + \sigma_{\rm mag}^2},\\
\hat{\bm{s}}_{t_k} &= \hat{\bm{s}}^{-}_{t_k} + \bm{k}_{t_k}\big[z_{t_k} - B_{\rm anom}(\hat{\bm{s}}^{-}_{t_k})\big],\\
\bm{\Sigma}_{t_k} &= \big(\bm{I}_2 - \bm{k}_{t_k}\,\bm{g}_{t_k}^{\rm T}\big)\,\bm{\Sigma}^{-}_{t_k},
\end{aligned}
\label{eq:ekf_update}
\end{equation}
where $\bm{k}_{t_k}$ is the Kalman gain and $\bm{g}_{t_k} = \nabla B_{\rm anom}(\hat{\bm{s}}^{-}_{t_k})$ is the map gradient, evaluated by central differences with a step of half a cell. Between map nodes, $B_{\rm anom}$ is interpolated bilinearly. The EKF is initialised at the INS position $\hat{\bm{s}}^{\rm INS}_{t_1}$ with $\bm{\Sigma}^{-}_{t_1} = (n_{\rm half}\,\Delta_{\rm cell})^2\,\bm{I}_2/3$, the covariance of the uniform prior of Equation~(\ref{Eq:Prior_Magnav}) over the first candidate grid, and is run over the whole flight without segmentation.

\subsection{Particle filter}\label{App:PF}

The bootstrap particle filter \citep{Gordon_1993} approximates the filtering density $p(\bm{s}_{t_k} \mid z_{t_1:t_k})$ by $N_{\rm p}$ weighted samples (particles), with $N_{\rm p} = 10^4$ and $10^5$ in Section~\ref{SubSec:Comparison}. The particles are initialised uniformly over the first candidate grid, consistent with Equation~(\ref{Eq:Prior_Magnav}). At each timestep, they are propagated by random draws from the motion model of Equation~(\ref{eq:motion_model}) and reweighted by the measurement factor of Equation~(\ref{eq:meas_magnav}), with $B_{\rm anom}$ interpolated bilinearly at each particle position. The position estimate is the weighted mean of the particles. The particles are resampled systematically \citep{Douc_2005} whenever the effective sample size $N_{\rm eff} = (\sum_{i=1}^{N_{\rm p}} w_i^2)^{-1}$, where $w_i$ denotes the normalised weight of the $i$-th particle, falls below $N_{\rm p}/2$. The particle filter is run over the whole flight without segmentation.

\subsection{Viterbi path}\label{App:ViterbiPath}

The Viterbi algorithm \citep{Viterbi_1967,Rabiner_1989} returns the most probable complete sequence $\hat{s}^{\rm Viterbi}_{\mathcal T}$ of Section~\ref{SubSec:point_estimation}. It operates on the same segmented factor graph as the forward-backward algorithm, with the sum in the pruned forward recursion of Equation~(\ref{eq:alpha_pruned}) replaced by a maximum and a traceback step that recovers the optimal path \citep[Section~III.B]{Rabiner_1989}. The max-product recursion reads
\begin{equation}
\delta_{t_k}[j] = f_m\left(z_{t_k},\,\bm{c}_{t_k}^j\right)
\max_{i \in \mathcal{A}_{t_{k-1}}}\left[\delta_{t_{k-1}}[i]\,
f_\tau\left(\bm{c}_{t_k}^j,\,\bm{c}_{t_{k-1}}^i\right)\right],
\quad j \in \mathcal{A}_{t_k},
\label{eq:viterbi_rec}
\end{equation}
where $\delta_{t_k}[j]$ is proportional to the joint probability of the measurements $z_{t_1:t_k}$ and the most probable partial sequence ending at candidate $\bm{c}^j_{t_k}$, and $\delta_{t_1}[j] = f_0(\bm{c}^j_{t_1})\,f_m(z_{t_1}, \bm{c}^j_{t_1})$. The Viterbi path is computed on the candidate grids and active sets of the factor graph. In the first segment, $f_0$ is the uniform prior of Equation~(\ref{Eq:Prior_Magnav}). In each subsequent segment, the last cell of the preceding path is propagated through one transition factor, which takes the place of the posterior carry-over of Equation~(\ref{eq:prior_carryover}).

Like the marginal MAP estimate of Equation~(\ref{eq:map}), the Viterbi path is restricted to cell centres, because it is computed without the sub-cell refinement of \citet[Section~3.3]{Li_2024}. The marginal MAP estimate is read off the same posterior $\gamma_{t_k}$ as the MMSE estimate, so the difference between their errors bounds the cost of quantisation (Section~\ref{SubSec:Idealisations}). We defer the sub-cell refinement of the Viterbi path to future work.

\section{Active set pruning threshold selection}\label{App:ActiveSetPruning}
The active-set pruning of Section~\ref{SubSec:Pruning} introduces the pruning threshold $\alpha_{\rm{prune}}$. At each timestep $t_k$, the active set $\mathcal{A}_{t_k}$ of Equation~(\ref{eq:active_set}) retains only those candidates whose measurement factor $f_m$ is at least $\alpha_{\rm prune}$ times the maximum measurement factor over all candidates at that timestep. Larger values of $\alpha_{\rm prune}$ yield smaller active sets and reduce the run time $t_{\rm run}$ of the forward-backward inference of Section~\ref{SubSec:forward_backward}, but risk discarding candidates that carry non-negligible posterior mass. In this appendix, we quantify the foregoing trade-off on $10^2$ sensor-error realisations of the representative test case of Section~\ref{Sec:Validation}. Each realisation is run without pruning, which we denote by $\alpha_{\rm prune} = 0$, and with pruning at seven values of $\alpha_{\rm prune}$ between $10^{-4}$ and $10^{-1}$.

We compare each pruned run with the unpruned run of the same realisation through four quantities. (i) The exclusion rate $P_{\rm excl}(\alpha_{\rm prune})$ is the percentage of the $N_{\rm FG}$ timesteps of a flight at which the cell nearest the true position is excluded from $\mathcal{A}_{t_k}$. (ii) The entropy bias is $\Delta H(\alpha_{\rm prune}) = N_{\rm FG}^{-1} \sum_{k=1}^{N_{\rm FG}} \big[H_{t_k}(\alpha_{\rm prune}) - H_{t_k}(0)\big]$, i.e.\ the change in the flight-mean entropy caused by pruning. (iii) The change in error is $\Delta e_{\rm pos}(\alpha_{\rm prune}) = e_{\rm med}(\alpha_{\rm prune}) - e_{\rm med}(0)$, where $e_{\rm med}$ is the per-trial median position error of Section~\ref{SubSec:Noise}. Positive values indicate that pruning increases the position error. (iv) The run time $t_{\rm run}(\alpha_{\rm prune})$ is the time taken to process one flight.

\begin{figure}
\centering
\includegraphics[width=1.0\textwidth]{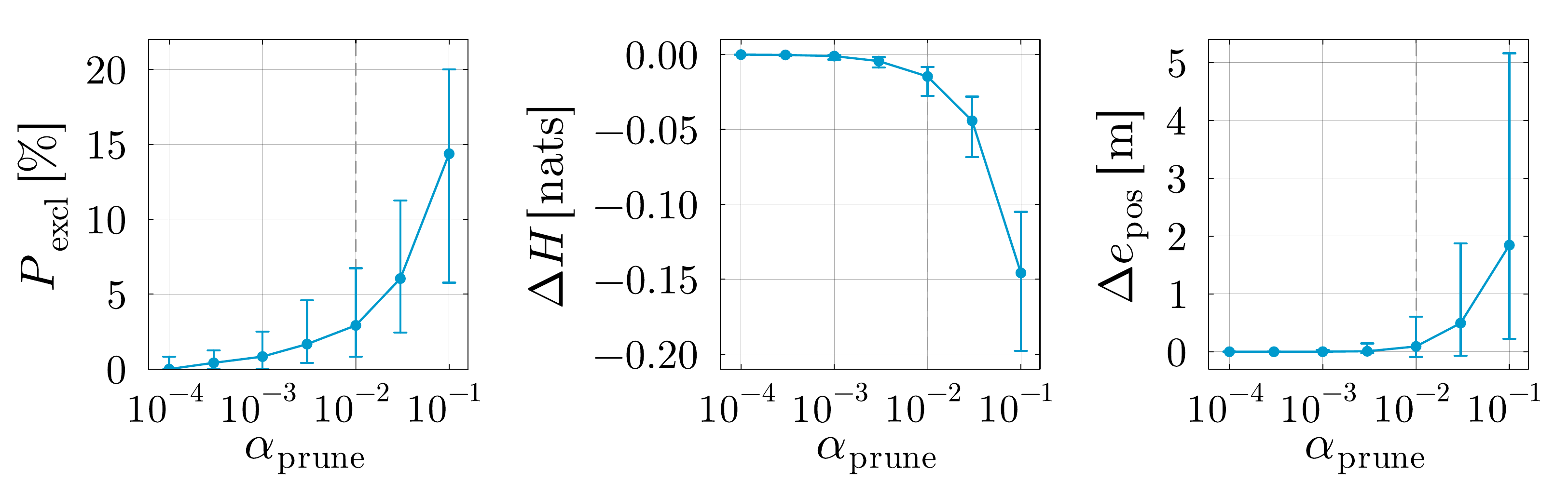}
\caption{Effect of the pruning threshold $\alpha_{\rm prune}$ on $10^2$ sensor-error realisations of the representative test case of Section~\ref{Sec:Validation}. Left panel: exclusion rate $P_{\rm excl}$, the percentage of timesteps at which the cell nearest the true position is excluded from the active set $\mathcal{A}_{t_k}$. Middle panel: entropy bias $\Delta H$ (units: nats), the change in the flight-mean entropy relative to the unpruned run. Right panel: change in the median position error $\Delta e_{\rm pos}$ (units: m) relative to the unpruned run. Points mark the median over the realisations, and bars span the 16th to 84th percentiles. The dashed vertical line marks $\alpha_{\rm prune} = 10^{-2}$, the setting adopted in this paper. The horizontal axes are plotted on a logarithmic scale}
\label{Fig:Pruned}
\end{figure}

In Figure~\ref{Fig:Pruned}, we plot $P_{\rm excl}$ (left panel), $\Delta H$ (middle panel), and $\Delta e_{\rm pos}$ (right panel) versus $\alpha_{\rm prune}$. Points mark the median over the $10^2$ realisations, and bars span the 16th to 84th percentiles. The dashed vertical line marks the setting $\alpha_{\rm prune} = 10^{-2}$ adopted in this paper (Table~\ref{tab:fg_params}).

Figure~\ref{Fig:Pruned} has three key features. First, $P_{\rm excl}$ rises with $\alpha_{\rm prune}$. Its median is $0.83\%$ at $\alpha_{\rm prune} = 10^{-3}$, $2.9\%$ at $\alpha_{\rm prune} = 10^{-2}$, and $14\%$ at $\alpha_{\rm prune} = 10^{-1}$. The active set is rebuilt at every timestep, so an exclusion removes the cell nearest the true position at that timestep only. Second, $\Delta H$ is negative, because the pruned cells carry zero posterior mass (Section~\ref{SubSec:point_estimation}). The median of $\Delta H$ is $-0.015$ nats at $\alpha_{\rm prune} = 10^{-2}$. For comparison, the median flight-mean entropy without pruning is $2.81$ nats. At $\alpha_{\rm prune} = 10^{-1}$, the median of $\Delta H$ is $-0.15$ nats. Third, $\Delta e_{\rm pos}$ increases with $\alpha_{\rm prune}$. Its median is below $0.01$~m for $\alpha_{\rm prune} \leq 10^{-3}$, $0.09$~m at $\alpha_{\rm prune} = 10^{-2}$, and $1.8$~m at $\alpha_{\rm prune} = 10^{-1}$.

Without pruning, the median run time of a flight is $t_{\rm run}(0) = 281$~s. Pruning reduces the median run time by a factor of $20$ at $\alpha_{\rm prune} = 10^{-4}$ and by a factor of $71$ at $\alpha_{\rm prune} = 10^{-1}$. At the setting $\alpha_{\rm prune} = 10^{-2}$ adopted in this paper, pruning reduces the median run time by a factor of $38$. The cost is an increase of $0.09$~m in the median position error, which is small relative to the map grid spacing $\Delta_{\rm cell} \approx 29.3$~m, and an entropy bias of $-0.015$~nats.

\section{Factor graph parameter estimation}\label{App:Sigma}

The factor graph parameters $\sigma_{\rm mag}$ and $\sigma_v$ employed in Sections~\ref{Sec:Validation} and~\ref{Sec:Performance} set the widths of the measurement and transition factors of Equations~(\ref{eq:meas_magnav}) and~(\ref{eq:trans_magnav}), respectively. Neither parameter is known in advance in an operational setting. The parameter $\sigma_{\rm mag}$ absorbs map interpolation error and residual compensation artefacts in addition to sensor noise (Section~\ref{SubSec:Meas}). The parameter $\sigma_v$ is an effective velocity uncertainty, which absorbs the time-correlated INS velocity errors omitted by the motion model of Equation~(\ref{eq:motion_model}) (Section~\ref{SubSec:Idealisations}). In addition, the magnetic measurements carry a bias, which is removed against the map before they enter the measurement factor. In Section~\ref{SubSec:DataPrep}, the measurements are debiased against the map along a reference path in successive passes [Equation~(\ref{eq:debias})], and the noise parameters are estimated by maximising the Bayesian evidence [Equation~(\ref{eq:evidence_max})]. In this appendix, we specify the procedure in detail. The debiasing passes, the noise parameters used within each pass, and the stopping rule are described in Appendix~\ref{App:SigmaPasses}. The grid on which the evidence is maximised is specified in Appendix~\ref{App:SigmaGrid}. The sensitivity of the position error to $\sigma_{\rm mag}$ and $\sigma_v$ on the representative test case of Section~\ref{Sec:Validation} is quantified in Appendix~\ref{App:SigmaSweep}.

\subsection{Debiasing passes}\label{App:SigmaPasses}

Equation~(\ref{eq:debias}) is applied in successive passes, because the map values along a reference path displaced from the true trajectory differ from those along the true trajectory (Section~\ref{SubSec:DataPrep}). Each pass runs the factor graph with provisional values of $\sigma_{\rm mag}$ and $\sigma_v$, which precede the evidence maximisation of Appendix~\ref{App:SigmaGrid}. On the first pass, the reference path is the INS position estimate, and the provisional values are $\sigma_{\rm mag} = 3$~nT and $\sigma_v = 3$~m\,s$^{-1}$, because the residuals along the INS path reflect the INS position error. On each subsequent pass, the reference path is the MMSE estimate of the preceding pass, i.e.\ $\bar{\bm{s}}_{t_k} = \hat{\bm{s}}^{\rm MMSE}_{t_k}$, and the provisional values are estimated from the residuals along it, viz.\
\begin{equation}
\sigma_{\rm mag} = 1.4826\,\operatorname{median}_k \big| r_{t_k} - \operatorname{median}_l\, r_{t_l} \big|, \qquad
\sigma_v = \frac{1}{\Delta t}\,\max_{a \in \{{\rm N}, {\rm E}\}}\, \operatorname{std}_k \big[ \bar{s}^{(a)}_{t_k} - \bar{s}^{(a)}_{t_{k-1}} - v^{(a)}_{t_{k-1}}\,\Delta t \big].
\label{eq:pass_sigmas}
\end{equation}
In Equation~(\ref{eq:pass_sigmas}), $r_{t_k} = z_{t_k} - B_{\rm anom}(\bar{\bm{s}}_{t_k})$ denotes the residual of the debiased measurement along the reference path, the factor 1.4826 scales the median absolute deviation to the standard deviation of a Gaussian distribution, and the displacements $\bar{s}^{(a)}_{t_k} - \bar{s}^{(a)}_{t_{k-1}}$ along the north ($a = {\rm N}$) and east ($a = {\rm E}$) axes are evaluated as in Equation~(\ref{eq:displacement_NE}). The passes stop when $\sigma_{\rm mag}$ and $\sigma_v$ each change by at most 5\% between successive passes, or after five passes.

\subsection{Evidence maximisation}\label{App:SigmaGrid}

To solve Equation~(\ref{eq:evidence_max}), we evaluate the factor-graph evidence of Equation~(\ref{eq:evidence_product}) on a two-dimensional grid of $5 \times 5$ points in $(\log_{10}\sigma_{\rm mag},\, \log_{10}\sigma_v)$, spaced by 0.2~dex, i.e.\ a factor of $10^{0.2} \approx 1.58$. The grid is centred on the provisional values of the last pass, which are computed along the MMSE estimate of the factor graph. If the maximum lies on the edge of the grid, the grid is recentred once on that maximum. The evidence is computed from the measurements debiased in the last pass, with the candidate grids of Section~\ref{SubSec:Discretisation} held at the centres of the last pass, so that every pair $(\sigma_{\rm mag}, \sigma_v)$ is evaluated on the same candidate states. The factor graph is then run with the evidence maximum $(\hat{\sigma}_{\rm mag}, \hat{\sigma}_v)$, on grids recentred as in Section~\ref{SubSec:Discretisation}. The final maximum lies on the edge of the grid in 3\% of the sensor-error realisations of Section~\ref{SubSec:Noise} and in none of the flight paths of Section~\ref{SubSec:Observability}.

\subsection{Sensitivity to the noise parameters}\label{App:SigmaSweep}

\begin{figure}
\centering
\includegraphics[width=1.0\textwidth]{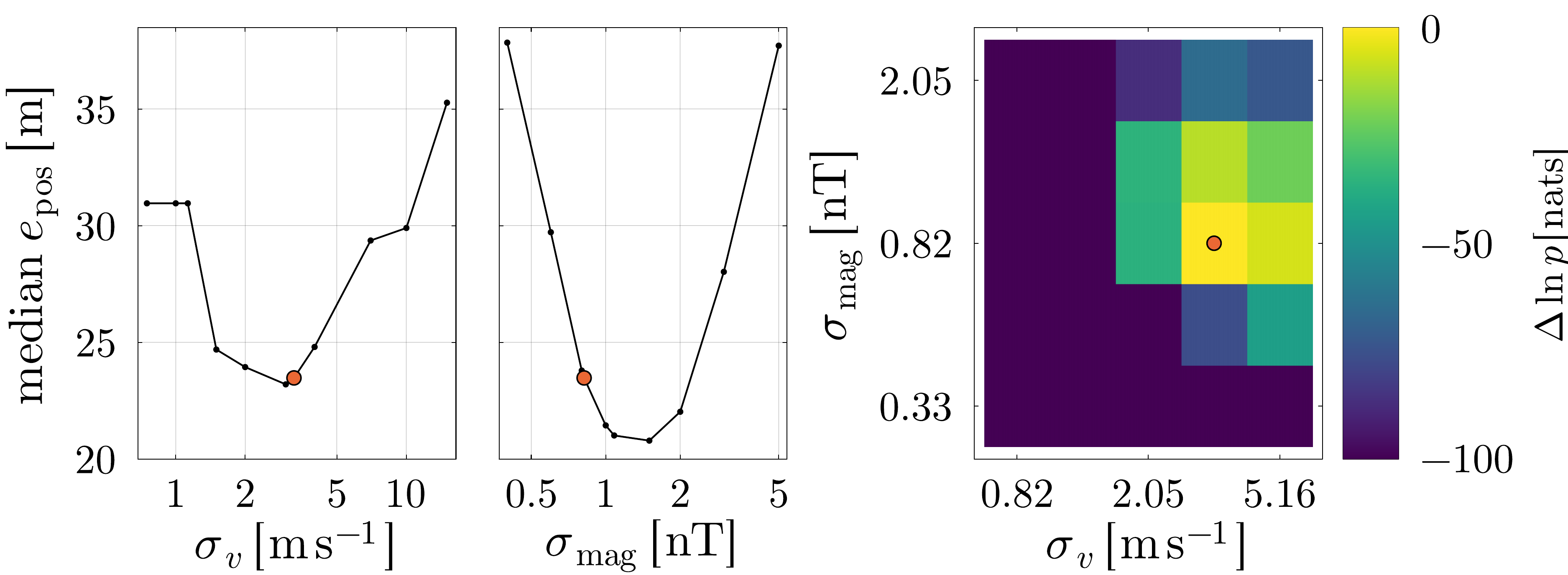}
\caption{Sensitivity of the median position error $e_{\rm pos}$ (units: m) to the noise parameters for the representative test case of Section~\ref{Sec:Validation}. Left panel: median error versus $\sigma_v$ (units: m\,s$^{-1}$) at $\hat{\sigma}_{\rm mag} = 0.82$~nT. Middle panel: median error versus $\sigma_{\rm mag}$ (units: nT) at $\hat{\sigma}_v = 3.26$~m\,s$^{-1}$. Right panel: logarithm of the evidence of Equation~(\ref{eq:evidence_product}) (units: nats) on the $5 \times 5$ grid of Appendix~\ref{App:SigmaGrid}, relative to its maximum; values below $-100$ nats are shown at $-100$ nats. The orange circle marks the evidence maximum in each panel. The left and middle panels share the vertical axis. The axes carrying $\sigma_v$ and $\sigma_{\rm mag}$ are plotted on a logarithmic scale}
\label{Fig:SigmaSweep}
\end{figure}

The procedure of Appendices~\ref{App:SigmaPasses} and~\ref{App:SigmaGrid} selects one pair $(\hat{\sigma}_{\rm mag}, \hat{\sigma}_v)$ per flight. To quantify how strongly the position error depends on this choice, we repeat the representative test case of Section~\ref{Sec:Validation} in two sweeps. The first varies $\sigma_v$ over $0.75 \leq \sigma_v/(1\,{\rm m\,s^{-1}}) \leq 15$ at $\hat{\sigma}_{\rm mag} = 0.82$~nT. The second varies $\sigma_{\rm mag}$ over $0.4 \leq \sigma_{\rm mag}/(1\,{\rm nT}) \leq 5$ at $\hat{\sigma}_v = 3.26$~m\,s$^{-1}$. In Figure~\ref{Fig:SigmaSweep}, we plot the median position error versus $\sigma_v$ (left panel) and versus $\sigma_{\rm mag}$ (middle panel), together with the log evidence on the $5 \times 5$ grid of Appendix~\ref{App:SigmaGrid}, relative to its maximum (right panel). The evidence maximum is marked in each panel.

Figure~\ref{Fig:SigmaSweep} has three key features. First, the median error varies by at most 3~m about the evidence maximum. In the $\sigma_v$ sweep, it lies between 23.2~m and 24.8~m for $2 \leq \sigma_v/(1\,{\rm m\,s^{-1}}) \leq 4$. In the $\sigma_{\rm mag}$ sweep, it lies between 20.8~m and 23.8~m for $0.8 \leq \sigma_{\rm mag}/(1\,{\rm nT}) \leq 1.5$. Outside these ranges, the median error increases. It reaches 35.3~m at $\sigma_v = 15$~m\,s$^{-1}$. It reaches 37.8~m at $\sigma_{\rm mag} = 0.4$~nT and 37.7~m at $\sigma_{\rm mag} = 5$~nT. For $\sigma_v \leq 1.13$~m\,s$^{-1}$, the median error is constant at 31.0~m, because the width $\sigma_v\,\Delta t$ of the transition factor is held at no less than $0.2\,\Delta_{\rm cell}$. Second, the median error at the evidence maximum lies close to the smallest median error of each sweep. It exceeds the smallest median error by 0.3~m in the $\sigma_v$ sweep and by 2.7~m in the $\sigma_{\rm mag}$ sweep. Third, the log evidence at each of the eight grid points adjacent to the maximum is lower by at least 6.3 nats.

\section{Magnetic map mismatch}\label{App:MapMismatch}

The measurements of Sections~\ref{Sec:Validation} and~\ref{Sec:Performance} are generated from the same magnetic anomaly map, interpolation, altitude, and upward continuation as the measurement factor of Equation~(\ref{eq:meas_magnav}). In practice, the map available to the navigator differs from the anomaly field along the flight path, because of survey and compilation errors, the upward continuation to the flight altitude, and errors in the assumed altitude. Let $B_{\rm true}(\bm{s})$ denote the anomaly field that generates the measurements. The map error, viz.\
\begin{equation}
\delta B(\bm{s}) = B_{\rm true}(\bm{s}) - B_{\rm anom}(\bm{s}),
\label{eq:map_error}
\end{equation}
enters every measurement of Equation~(\ref{eq:obs_model}). Because $\delta B$ varies slowly along the track, its values at consecutive timesteps are correlated. The measurement factor of Equation~(\ref{eq:meas_magnav}) does not model this correlation, because it assumes that the residuals at different timesteps are independent [Section~\ref{SubSec:Idealisations}, idealisation (ii)]. In this appendix, we quantify the effect of $\delta B$ on the first $10^2$ trials of Section~\ref{SubSec:Noise}, each paired with the run without mismatch, i.e.\ with the same INS and sensor-error realisations. The measurements are debiased and the noise parameters are estimated on each trial, as in Section~\ref{SubSec:DataPrep}. We summarise $\delta B$ along the true track by its root-mean-square value, $\delta B_{\rm rms} = [N_{\rm FG}^{-1} \sum_{k=1}^{N_{\rm FG}} \delta B(\bm{s}_{t_k})^2]^{1/2}$.

We consider two sources of mismatch: (i) the map and its continuation; and (ii) the altitude. In case (i), the measurements are generated from a second aeromagnetic anomaly map of the same area, \texttt{Renfrew\_555}, compiled at a survey altitude of 555~m and continued upward by 445~m to the flight altitude $h$. The factor graph retains \texttt{Renfrew\_395}, continued upward by 605~m. In case (ii), the measurements are generated at $h$, and the factor graph's map is continued to $\Delta h = 50$~m and 100~m above $h$, so that $\delta B \approx (\partial B_{\rm anom}/\partial h)\,\Delta h$ (Section~\ref{SubSec:SimIdealisations}). In Table~\ref{tab:map_mismatch}, we list $\delta B_{\rm rms}$, the median position error pooled over all timesteps and trials, and the change in the per-trial median error relative to the paired run without mismatch.

\begin{table}[t]
\centering
\caption{Magnetic map mismatch on the first $10^2$ trials of Section~\ref{SubSec:Noise}, each paired with the run without mismatch. $\delta B_{\rm rms}$ is the root-mean-square map error along the true track at the flight altitude. The median error is pooled over all timesteps and trials. The change is the per-trial median error minus that of the paired run: median over trials, with the 16th and 84th percentiles in brackets. The median INS error on these trials is 341~m.}
\label{tab:map_mismatch}
\begin{tabular}{@{}lccc@{}}
\toprule
{\bfseries Case} & $\bm{\delta B_{\rm rms}}$ {\bfseries (nT)} & {\bfseries Median error (m)} & {\bfseries Change (m)} \\
\midrule
No mismatch                                  & 0    & 33.2  & -- \\
Measurements from \texttt{Renfrew\_555}      & 0.11 & 33.4  & $0.0$ ($-1.5$, $+1.7$) \\
Map continued to $h + 50$ m                  & 1.96 & 80.9  & $+45.7$ ($+31.8$, $+69.9$) \\
Map continued to $h + 100$ m                 & 3.82 & 128.5 & $+90.9$ ($+73.7$, $+132.4$) \\
\bottomrule
\end{tabular}
\end{table}

Table~\ref{tab:map_mismatch} has two key features. First, in case (i), $\delta B_{\rm rms} = 0.11$~nT. The median error is 33.4~m, to be compared with 33.2~m without mismatch. The median paired change is 0.0~m. Second, in case (ii), $\delta B_{\rm rms}$ is 1.96~nT for $\Delta h = 50$~m and 3.82~nT for $\Delta h = 100$~m. The median error rises to 80.9~m for $\Delta h = 50$~m and to 128.5~m for $\Delta h = 100$~m. In case (ii), the error increases on every trial and remains below the median INS error of 341~m. The lag-one autocorrelation of $\delta B$ at $\Delta t_{\rm FG} = 5$~s is 0.95 in case (i) and 0.93 at each altitude offset of case (ii).

The foregoing results delimit the conditions under which the white-noise measurement factor of Equation~(\ref{eq:meas_magnav}) holds. In case (i), $\delta B_{\rm rms} = 0.11$~nT is small compared with the median $\hat{\sigma}_{\rm mag} = 0.91$~nT of the trials of Section~\ref{SubSec:Noise}, and the median position error of the MMSE estimate is insensitive to $\delta B$. In case (ii), $\delta B_{\rm rms} \gtrsim \sigma_{\rm mag}$, and the median position error increases, because the measurement factor treats the correlated map error as white noise.

\section{Sliding segment window}\label{App:Sliding}

The non-overlapping segmentation of Section~\ref{SubSec:Segmentation} partitions the flight into $M$ segments $\mathcal{T}^{(1)}, \ldots, \mathcal{T}^{(M)}$, each of length $T$ timesteps. Consecutive segments are connected by the prior carry-over of Equation~(\ref{eq:prior_carryover}). Within each segment, timesteps near the segment boundaries are conditioned on fewer measurements than those near the segment centre. The first timestep draws on $T-1$ later measurements via the backward sweep, but on no earlier measurements within the segment. The last timestep draws on $T-1$ earlier measurements, but on no later ones. At the last timestep, $\beta_{t_T} = 1$ [Equation~(\ref{eq:beta_init})], and the factor graph returns the PMF estimate (Section~\ref{SubSec:MagNavError}). For applications in which every position estimate should draw on later measurements, e.g.\ post-flight trajectory reconstruction, where the delay of the backward sweep and the computational cost are secondary, an alternative strategy is a sliding segment window, whose segments overlap. In this appendix, we specify the sliding segment window, which generalises the segmentation of Section~\ref{SubSec:Segmentation}, and quote its performance on the representative test case of Section~\ref{Sec:Validation}.

Fix the segment length $T$ and introduce a stride $S < T$, so that consecutive segments overlap by $T - S$ timesteps. The sliding segments are indexed by $m' \in \{1, \ldots, M'\}$, where $M' = \lfloor (N_{\rm FG} - T) / S \rfloor + 1$, and each segment covers timesteps
\begin{equation}
    \mathcal{T}^{(m')} = \left\{\left(m'-1\right)\,S + 1,\;\ldots,\;\left(m'-1\right)\,S + T\right\}.
    \label{eq:sliding_segment}
\end{equation}
When $S = T$ (no overlap), the scheme reduces to the non-overlapping partition of Section~\ref{SubSec:Segmentation}. When $S < T$, the number of segments, and hence the computational cost, increases by a factor of approximately $T/S$ relative to the non-overlapping case.

Each sliding segment is processed by the forward-backward algorithm of Section~\ref{SubSec:forward_backward}. The prior of segment $m'$ is propagated from the $S$-th timestep of segment $m'-1$, i.e.\ the last timestep preceding the start of segment $m'$, through one application of the transition factor, viz.\
\begin{equation}
f_0\left(\bm{c}_{t_1^{(m')}}^j\right)
= \sum_{i=1}^{N_s}
  \frac{\alpha_{t_S^{(m'-1)}}[i]}{\sum_{\ell=1}^{N_s} \alpha_{t_S^{(m'-1)}}[\ell]}\;
  f_\tau\!\left(\bm{c}_{t_1^{(m')}}^j,\;
                \bm{c}_{t_S^{(m'-1)}}^i\right),
\label{eq:sliding_carryover}
\end{equation}
followed by normalisation. Equation~(\ref{eq:sliding_carryover}) carries the normalised forward variable, in place of the marginal posterior $\gamma$ of Equation~(\ref{eq:prior_carryover}), because $\gamma_{t_S^{(m'-1)}}$ is conditioned on the measurements at timesteps $S+1, \ldots, T$ of segment $m'-1$, which segment $m'$ processes again. The normalised forward variable is conditioned on the measurements up to $t_S^{(m'-1)}$ only, so each measurement enters the inference once.

Because segments overlap, each interior timestep $t$ is covered by multiple segments. The MMSE estimate at $t$ is taken from the segment $m^*(t)$ whose temporal centre is closest to $t$,
\begin{equation}
    m^*(t) = \arg\min_{m' :\, t \in \mathcal{T}^{(m')}}
              \bigl| t - \bar{t}^{(m')} \bigr|,
    \label{eq:sliding_merge}
\end{equation}
where $\bar{t}^{(m')} = (m'-1)\,S + (T+1)/2$ is the temporal midpoint of segment $m'$. Accordingly, every estimate before the midpoint of the last segment draws on at least $(T-S)/2$ later measurements within its segment.

We apply the sliding segment window with $S = 20$, i.e.\ 50\% overlap and $M' = 11$ segments, to the representative test case of Section~\ref{Sec:Validation}, with the factor graph parameters of Table~\ref{tab:fg_params}. Relative to the non-overlapping segmentation ($S = T = 40$, $M = 6$), the median position error falls from 23.5~m to 22.5~m. The largest error is 82.2~m, to be compared with 82.7~m for the non-overlapping segmentation. The computational cost increases by a factor of approximately $T/S = 2$.

The results of Sections~\ref{Sec:Validation} and~\ref{Sec:Performance} use the non-overlapping segmentation of Section~\ref{SubSec:Segmentation}. The evaluation of the sliding segment window on the ensembles of Section~\ref{Sec:Performance} is deferred to future work.

\section{Fisher information for scalar field positioning}\label{App:FIM}

The Fisher information matrix (FIM) quantifies the sensitivity of a likelihood function to parameters of interest and, through the Cram\'{e}r--Rao bound, provides a lower bound on the variance of any unbiased estimator \citep{Kay_1993,VanTrees_2004}. In the MagNav context,
the FIM characterises the position information available from scalar magnetic anomaly observations $B_{\rm anom}(\phi, \lambda)$ and motivates the observability metrics
employed in Sections~\ref{SubSec:ShannonEntropy}
and~\ref{SubSec:Observability}. We derive the per-observation and
per-segment FIM below for the observation model of
Section~\ref{SubSec:Meas}.

Consider the observation model of Equation~(\ref{eq:obs_model}), in which a
scalar magnetic anomaly measurement $z_{t_k}$ is related to the aircraft
position $\bm{s}_{t_k} = (\phi_{t_k}, \lambda_{t_k})$ by
$z_{t_k} = B_{\rm anom}(\bm{s}_{t_k}) + \epsilon_{t_k}$, with
$\epsilon_{t_k} \sim \mathcal{N}(0,\,\sigma_{\rm mag}^2)$. The conditional
log-likelihood of a single observation is
\begin{equation}
\ln p\left(z_{t_k} \mid \bm{s}_{t_k}\right)
= -\frac{\left[z_{t_k} - B_{\rm anom}\left(\bm{s}_{t_k}\right)\right]^2}{2\,\sigma_{\rm mag}^2}
  - \frac{1}{2}\ln\left(2\pi\sigma_{\rm mag}^2\right).
\label{eq:loglik_scalar}
\end{equation}
The FIM is defined as
the expected negative Hessian of the log-likelihood, Equation~(\ref{eq:loglik_scalar}), with respect to the
position coordinates \citep{Kay_1993,VanTrees_2004}, viz.\
\begin{equation}
J_{t_k}
= -\mathbb{E}\!\left[
    \frac{\partial^2 \ln p\left(z_{t_k} \mid \bm{s}_{t_k}\right)}
         {\partial \bm{s}_{t_k}\,\partial \bm{s}_{t_k}^{\rm T}}
  \right]
= \frac{1}{\sigma_{\rm mag}^2}\,
  \nabla B_{\rm anom}\left(\bm{s}_{t_k}\right)\,\left[\nabla B_{\rm anom}\left(\bm{s}_{t_k}\right)\right]^{\rm{T}}.
\label{eq:fim_scalar}
\end{equation}
Because $J_{t_k}$ is the outer product of
a single vector, it is rank-1 with one non-zero eigenvalue
$\lambda_{\max} = \|\nabla B_{\rm anom}\|^2/\sigma_{\rm mag}^2$ aligned with $\nabla B_{\rm anom}$ and a zero eigenvalue in the orthogonal direction. A single scalar observation therefore provides no information about position along the contours of the anomaly field.

For $T$ independent observations along a trajectory segment, the accumulated FIM is
\begin{equation}
J^{(m)} = \sum_{t \in \mathcal{T}^{(m)}} J_t
    = \frac{1}{\sigma_{\rm mag}^2}
      \sum_{t \in \mathcal{T}^{(m)}}
      \nabla B_{\rm anom}\left(\bm{s}_t\right)\,\left[\nabla B_{\rm anom}\left(\bm{s}_t\right)\right]^{\rm{T}},
\label{eq:fim_accumulated}
\end{equation}
where $\mathcal{T}^{(m)}$ denotes the timesteps of segment $m$ (Section~\ref{SubSec:Segmentation}).
If the aircraft trajectory samples sufficiently diverse gradient directions,
$J^{(m)}$ becomes full-rank and the minimum eigenvalue
$\lambda_{\min}[J^{(m)}] > 0$ characterises the least informative spatial
direction over the segment. The Cram\'{e}r--Rao bound states that the
covariance of any unbiased position estimator satisfies
$\mathrm{Cov}(\hat{\bm{s}}) \succeq [J^{(m)}]^{-1}$ \citep{Kay_1993,VanTrees_2004}, so that  $\{\lambda_{\min}[J^{(m)}]\}^{-1/2}$ is a lower bound on the position error standard deviation in the worst-case direction. The scalar cumulative Fisher information $J(T) \approx T\,\overline{\|\nabla B_{\rm anom}\|^2}\,/\,\sigma_{\rm mag}^2$ employed in the segment-length selection of Section~\ref{SubSec:SelectingT} equals the trace of $J^{(m)}$ when the per-timestep gradient magnitude is replaced by its path-averaged value. This provides an orientation-averaged information measure. By contrast, $\lambda_{\min}[J^{(m)}]$ measures the information accumulated in the least-informative horizontal direction, and is therefore the segment-level observability metric used in Sections~\ref{Sec:Validation} and~\ref{Sec:Performance}. The Fisher-information calculation is used here as a local map-observability diagnostic rather than a complete prediction of estimator error, because the factor-graph posterior may be discrete, multimodal, affected by active-set pruning, and influenced by transition-model mismatch.

\end{appendices}

\bibliographystyle{apalike}
\bibliography{main}

\end{document}